\documentclass{jpp}
\usepackage{graphicx}
\usepackage{natbib}
\usepackage{color}

\ifCUPmtlplainloaded \else
  \checkfont{eurm10}
  \iffontfound
    \IfFileExists{upmath.sty}
      {\typeout{^^JFound AMS Euler Roman fonts on the system,
                   using the 'upmath' package.^^J}%
       \usepackage{upmath}}
      {\typeout{^^JFound AMS Euler Roman fonts on the system, but you
                   dont seem to have the}%
       \typeout{'upmath' package installed. JPP.cls can take advantage
                 of these fonts, if you use 'upmath' package.^^J}%
      }
  \else
  \fi
\fi

\ifCUPmtlplainloaded \else
  \checkfont{msam10}
  \iffontfound
    \IfFileExists{amssymb.sty}
      {\typeout{^^JFound AMS Symbol fonts on the system, using the
                'amssymb' package.^^J}%
       \usepackage{amssymb}%
       \let\le=\leqslant  
       \let\ge=\geqslant  
      }{}
  \fi
\fi

\ifCUPmtlplainloaded \else
  \IfFileExists{amsbsy.sty}
    {\typeout{^^JFound the 'amsbsy' package on the system, using it.^^J}%
     \usepackage{amsbsy}}
    {}
\fi

\def\ADD#1{{\textcolor{black}{#1}}}
\def\eg{{\it e.g.}\ } 
\def\ie{{\it i.e.}\ }

\def\kk{{\bf k}}
\def\kpa{k_\parallel}
\def\kpe{k_\perp}
\def\bb{{\bf b}}
\def\uu{{\bf u}}

\def\bell{{\pmbmath{\ell}}}
\def\be{\begin{equation}}
\def\ee{\end{equation}}
\def\ba{\begin{eqnarray}}
\def\ea{\end{eqnarray}}
\def \pmbmath{\mathpalette\pmbmathaux}
\def \pmbmathaux#1#2{
         \pmbtext{$#1#2$}}
\def \pmbtext#1{\leavevmode
     \setbox0\hbox{#1}
     \kern0,4pt \copy0 \kern-\wd0
     \kern-0,2pt \raise0,3pt \box0 }

\def\kkp{{\bf k}_{\perp}}
\def\pp{{\bf p}}
\def\qq{{\bf q}}

\def\diso{d\ppn d\qpn d\ppa d\qpa}
\def\disot{d\tppn d\tqpn d\tppa d\tqpa}
\def\kpn{k_{\perp}}
\def\ppn{p_{\perp}}
\def\qpn{q_{\perp}}
\def\kpa{k_{\parallel}}
\def\ppa{p_{\parallel}}
\def\qpa{q_{\parallel}}
\def\tppn{\tilde p_{\perp}}
\def\tqpn{\tilde q_{\perp}}
\def\tppa{\tilde p_{\parallel}}
\def\tqpa{\tilde q_{\parallel}}

\def\nnn{{\tilde n}}
\def\mmm{{\tilde m}}
\def\xx{{\bf x}}
\def\RR{{\bf R}}
\def\HH{{\cal H}}
\def\aa{{\bf a}}
\def\AA{{\bf A}}

\title[Entanglement of helicity and energy in KAW/whistler turbulence]
{Entanglement of helicity and energy in kinetic Alfv\'en wave/whistler turbulence}

\author[S. Galtier and R. Meyrand]%
{S\ls \'E\ls B\ls A\ls S\ls T\ls I\ls E\ls N\ns G\ls A\ls L\ls T\ls I\ls E\ls R
\thanks{sebastien.galtier@lpp.polytechnique.fr} \and
R\ls O\ls M\ls A\ls I\ls N\ns M\ls E\ls Y\ls R\ls A\ls N\ls D}
\affiliation{Laboratoire de Physique des Plasmas, Ecole Polytechnique, F-91128 Palaiseau Cedex, France}

\date{\today}
\begin{document}

\maketitle

\begin{abstract}
The role of magnetic helicity is investigated in kinetic Alfv\'en wave and oblique whistler turbulence in presence of a relatively 
intense external magnetic field $b_0 {\bf e_\parallel}$. In this situation, turbulence is strongly anisotropic and the fluid equations 
describing both regimes are the reduced electron magnetohydrodynamics (REMHD) whose derivation, originally made from the 
gyrokinetic theory, is also obtained here from compressible Hall MHD. 
\ADD{We use the asymptotic equations derived by Galtier \& Bhattacharjee (2003) to study the REMHD dynamics in the weak 
turbulence regime.}
The analysis is focused on the magnetic helicity equation for which 
we obtain the exact solutions: they correspond to the entanglement relation, $n+\tilde n = -6$, where $n$ and $\tilde n$ are 
the power law indices of the perpendicular (to ${\bf b_0}$) wave number magnetic energy and helicity spectra respectively. 
Therefore, the spectra derived in the past from the energy equation only, namely $n=-2.5$ and $\tilde n = - 3.5$, are not the 
unique solutions to this problem but rather characterize the direct energy cascade. The solution $\tilde n = -3$ is a limit imposed 
by the locality condition; it is also the constant helicity flux solution obtained heuristically. The results obtained offer a new 
paradigm to understand solar wind turbulence at sub-ion scales where it is often observed that $-3 < n < -2.5$. 
\end{abstract}


%%%%%%%%%%%%%%%%%%%%%%%%
\section{Introduction}
\begin{figure}
\centerline{\includegraphics[height=7cm,width=10cm]{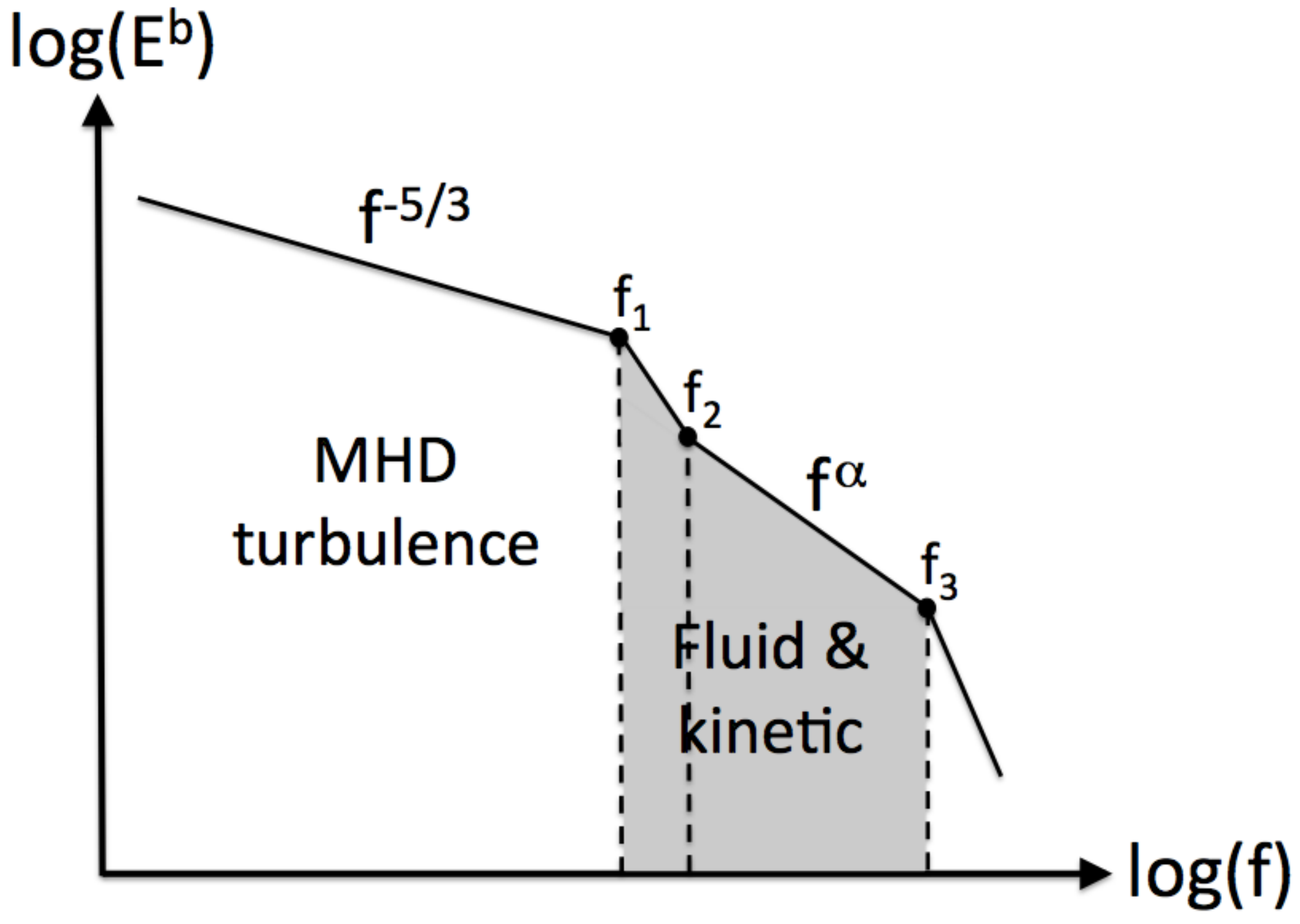}}
\caption{\ADD{Schematic view of the magnetic fluctuation spectrum observed in the solar wind at one astronomical unit. 
Three breaks are shown at frequencies $f_1$, $f_2$ and $f_3$. The large scale ($f<f_1 \sim 0.5$Hz) spectrum, characterized 
by a narrow range of power law indices with a peak near $-5/3$ \citep{smith06}, is generally interpreted as an MHD turbulence 
cascade. At sub-ion scales ($f> f_1$) the physical 
properties change drastically. After a stiff transition between breaks $f_1$ and $f_2$ \citep{sahraoui10}, a $f^{\alpha}$ spectrum 
is clearly observed such that $\alpha \in [-3.1,-2.5]$ with a value often around $-2.8$ \citep{fouad2013}. At sub-electron scales 
($f>f_3 \sim 50$Hz) another stiff variation -- possibly in power law -- is measured but the instrumental noise level is quickly 
reached. The gray part is the frequency domain where both kinetic Alfv\'en waves and whistler waves are detected and where 
both fluid and kinetic models are used.}}
\label{Fig0}
\end{figure}

Fluctuations in space plasmas exhibit a multitude of time and length scales such as the ion or electron cyclotron frequencies, 
inertial lengths and Larmor radii. For example, in the solar wind where {\it in situ} measurements are accessible the turbulent 
velocity, magnetic and density fluctuation spectra are characterized by extended power laws observed in the frequency range 
$10^{-5}$\,Hz $\le f \le 10^{2}$\,Hz \citep{bill82,Goldstein99,galtier06a,carbone12}. As expected, when one probes the solar wind 
plasma towards high frequencies, the physical properties evolve and several breaks in the magnetic field fluctuation spectrum 
are detected \citep{smith06,alexandrova09,sahraoui10}. 
\ADD{In figure\,\ref{Fig0} a schematic view of the magnetic field fluctuation spectrum observed in the solar wind is reported.}
For example, the break detected at 
the frequency $f_1 \sim 0.5$\,Hz is attributed to the decoupling between ions and electrons and defines, therefore, the scale 
at which one has to abandon the standard magnetohydrodynamic (MHD) model. The precise mechanism that drives 
the physics is still unclear: basically, the frequency range $f_1 \le f \le f_3$ is seen as either a dissipative range or/and a new 
turbulence regime. The difficulty resides, in particular, in the collisionless nature of the plasma and also its anisotropy. 
\ADD{Note that typical solar wind plasma parameters at one astronomical unit are: $\beta_i \sim \beta_e \sim 1$ (ratio of thermal 
to magnetic pressure for ion $\beta_i$ and electron $\beta_e$), $T_i / T_e \sim 1$ (ratio of ion to electron temperature), 
$d_i \sim 100$\,km (ion inertial length) and $f_{ci} \sim 0.1$\,Hz (ion cyclotron frequency).}

Both kinetic and fluid models are used to investigate the difficult problem of plasma turbulence at sub-ion scales 
\citep{Ghosh97,sheko09,meyrand10,rudakov}. For example, for the solar wind we often evoke kinetic Alfv\'en waves (KAW) and 
whistler waves \citep{sahraoui12} \ADD{which can be both excited for scales smaller than $d_i$.} 
The main difference between KAW and whistlers is in the dynamics of ions which rapidly adjust to the fluctuating electric potential 
in the former case, whereas they are dynamically irrelevant in the latter. 
\ADD{By definition, classical fluid models (\eg Hall MHD or electron MHD) are not able to catch kinetic effects and their 
use at sub-ion scales is mainly relevant for the investigation of the turbulent dynamics. In the solar wind case it is 
believed that the origin of the power law spectra for $f>f_2$ (see figure\,\ref{Fig0}) could be attributed mainly to turbulence 
which would imply that kinetic effects are irrelevant to understand the {\it statistical} properties of the magnetic 
fluctuations \citep{bill14}. Following this remark, we shall recall below the limit of validity of the fluid models.}
In the fluid case, the electron momentum equation with all electron inertia terms neglected gives the generalized ideal Ohm's 
law (in SI units):
\be
{\bf E} + \uu_e \times \bb - {\nabla p_e \over n_e e} = 0 \, , \label{gohm}
\ee 
where ${\bf E}$ is the electric field, $\uu_e$ the electron velocity, $\bb$ the magnetic field, $n_e$ the electron density, $e$ the 
magnitude of the electron charge, and $p_e$ the electron pressure. 
The introduction of equation (\ref{gohm}) into the Maxwell--Faraday's law, with the MHD momentum equation, and for example 
the polytropic closure, lead to the so-called Hall MHD system in which the Hall effect becomes dominant at length scales 
smaller than the ion inertial length $d_i$ ($d_i \equiv c / \omega_{pi}$ with $c$ the speed of light and $\omega_{pi}$ the ion 
plasma frequency) and time-scales of the order, or shorter than, the ion cyclotron period $\omega_{ci}^{-1}$. 
Note that in Hall MHD, the electron pressure $p_e$ is assumed to be a scalar (this can be justified in the collisional 
limit or in the isothermal electron fluid approximation \citep{sheko09}). 
\ADD{The limit of validity of Hall MHD may be discussed at the level of the dispersion relation. 
(We focus the discussion on frequencies where the electron inertia is not felt.) As shown by \cite{hirose}, 
the Hall MHD dispersion relation is a rigorous limit of Vlasov-Maxwell kinetic theory in the limit of cold ions, \ie $T_i = 0$. 
Under this limit, the ion Landau resonance becomes negligible which explains why physically the fluid model may be relevant. 
However, two important comments have to made here. First, this limitation was discussed quantitatively by \cite{howes06} 
(see also \cite{sahraoui12}) 
from the numerical resolution of the dispersion relations. It was found that even at $T_i = T_e$ (with $\beta =1$)  
the parallel whistler, oblique 
whistler and kinetic Alfv\'en waves are well described by Hall MHD whereas for example the slow mode represents an 
unphysical/spurious wave that does not exist in a weakly collisional plasma. Second, the simple analytical demonstration 
made on the dispersion relation does not say anything about the validity of Hall MHD in the full turbulent regime for which 
the statistical contribution of kinetic effects is still not well documented. If this contribution is negligible \citep{bill14} or if the 
three previous waves are not affected by the spurious waves, then Hall MHD may be a relevant model for plasmas even when 
$T_i \sim T_e$ (and $\beta \sim 1$). An example is given with incompressible ($\beta \gg 1$) Hall MHD in the regime of wave 
turbulence. As shown by \cite{galtier06}, it is possible to get spectral predictions only for the right circularly polarized wave 
(so without the feedback of the left circularly polarized wave) which may describe parallel whistler, oblique whistler or KAW 
(see the discussion in section \ref{sect2}). Also, it is though that as long as a fluid model like Hall MHD is able to 
predict statistical properties compatible with the observations the pure turbulent cascade scenario has to be considered as a 
central mechanism to transfer energy scale by scale until the electron scales. Beyond the solar wind,}
Hall MHD is widely used to investigate several questions such as the origin of the fast magnetic reconnection 
\citep{Bhatt04,shepherd} or the formation and disruption of Alfv\'enic filaments \citep{dreher}. 

Three-dimensional Hall MHD turbulence is much more difficult to investigate numerically than pure MHD because the Hall 
effect brings a new kind of nonlinear term with a second-order derivative which sometimes forces us to use lower-dimensional 
models \citep{Ghosh96,galtier07}. Because of this difficulty, it is interesting to investigate first the incompressible 
limit for which the Hall effect is asymptotically large, \ie the so-called electron MHD regime \citep{kingsep,goldreich92,das,diamond11}. 
In this limit, the ions can be considered as a motionless neutralizing background such that the electron flow determines entirely 
the electric current. Since the ions are static, electron compressibility corresponds to a violation of quasineutrality and reciprocally, 
therefore, electron MHD is only valid for large enough $\beta_e$ \citep{biskamp00}. This is basically the reason why one can 
recover electron MHD from incompressible Hall MHD. Electron MHD plays important roles for example in laser plasmas 
\citep{sentoku,cai} and magnetic field reconnection \citep{bulanov,drake,mandt,das00,lukin}. 
Direct numerical simulations of isotropic electron MHD show that the turbulent magnetic energy spectrum scales like $k^{-7/3}$ 
\citep{biskamp96,ng03}. This scaling was explained by a heuristic model \`a la Kolmogorov which turns out to be compatible 
dimensionally with an exact relation derived for third-order correlation functions \citep{galtier08}. 
It is widely believed that Hall MHD should exhibit the same (magnetic) spectrum \ADD{as electron MHD because the latter 
is simply the $k d_i \gg 1$ limit of the former.}
However, in the framework of three-dimensional incompressible Hall MHD, a recent study has revealed the 
influence of the (left or right) polarity on the (magnetic) energy spectra \citep{meyrand12} with the possibility to get different 
power laws at different scales, a tendency already observed with low-dimensional \ADD{shell models} \citep{galtier07,hori,pandit}. 

The case of a plasma embedded in an external magnetic field $\bb_0$ is even more difficult to investigate because numerically it 
brings a strong constraint on the time step and physically it leads to anisotropy. Nevertheless, this situation has been investigated 
numerically and analytically in strong \citep{Ghosh97,mininni07} and weak \citep{galtier06,Sahraoui07} Hall MHD turbulence. 
In the electron MHD limit, three-dimensional direct numerical simulations reveal that a state of critical balance may be reached 
\citep{cho04} when \ADD{in particular} the external magnetic field $b_0$ is of the order of the fluctuations\footnote{\ADD{Critical 
balance conjectures that in the strong turbulence regime, we have a scale by scale balance between the linear propagation and 
nonlinear time scales. For electron MHD the balance relation reads: $b \kpe \sim b_0 \kpa$ \citep{cho04}. The weak turbulence 
regime corresponds to $b \kpe \ll b_0 \kpa$, a situation reached when for example $b_0$ is strong enough.}}.
\ADD{When $b_0$} is much larger than the fluctuations an anomalous scaling in $k_\perp^{-8/3}$ has been found for the perpendicular 
(to ${\bf b_0}$) wave number energy spectra \citep{meyrand13} which may be explained with a heuristic model \citep{Galtier05}. 

In addition to the large amount of research with fluid models, kinetic models are also widely used. A kinetic theory of plasma turbulence 
is, however, extremely difficult to reach because of the conceptual difficulty to manage \eg with the multidimensional phase space and 
the multitude of phenomena that are included. For these reasons, simplifications are generally made in order to catch the most 
interesting part of the nonlinear dynamics. For example, \ADD{most of} the gyrokinetic theory\ADD{/simulations} \citep{sheko09} 
assumes that the distribution is close to a Maxwellian which is a rather strong assumption for space plasmas 
\ADD{(note that, for simplicity, this assumption is also made for axisymmetric tokamaks in \citet{frieman82})}. 
Additionally, it is also assumed that the turbulent magnetic fluctuations are relatively 
small compared to the mean magnetic field, spatially anisotropic with respect to it and that their frequency is low compared to the ion 
cyclotron frequency. Under these hypotheses, it is possible to make numerical simulations and to follow \eg the nonlinear dynamics 
driven by the KAW \citep{howes08}. The Reynolds number -- or in other words the size of the inertial range -- is however still 
significantly limited compared to pure fluid simulations and it might question the relevance of the results obtained for space plasmas 
like solar wind turbulence. Fortunately, the KAW cascade may also be described by a simple system called reduced electron MHD 
(REMHD) \ADD{valid for any temperature ratio $T_i/T_e$ and $\beta_i$, and}
whose derivation assumes an ordering \ADD{for the different variables} which implies, in particular, that the density 
variations can only be relatively small \ADD{compared to equilibrium constant density.} The form of the REMHD equations is close 
to the well-known (incompressible) electron MHD equations and in the strongly anisotropic limit ($\kpe \gg \kpa$) they become even 
mathematically similar which means that the nonlinear dynamics of strongly oblique whistler and KAW are the same 
\citep{cho09,sheko09,boldyrev12}: \ADD{thus, electron MHD is not only the high-beta limit ($\beta_i \to +\infty$) of REMHD but 
mathematically it is also possible to show that both systems are rigorously equivalent $\forall \beta_i$ and $\forall \, T_i/T_e$
(see the discussion in section \ref{sect2}).} 
From this remark, we can conclude that the weak turbulence predictions \citep{voitenko98,galtier03b,GB05,galtier06} are the 
same for both waves since it is characterized by a strong anisotropy. 

The role of magnetic helicity -- the scalar product of the magnetic field with the magnetic vector potential -- on electron MHD  has 
been investigated experimentally  through the production of electron MHD heat pulses and the study of their transport 
\citep{stenzel95,stenzel96,rousculp97}. In turbulence, the first effect reported in the literature is from two-dimensional direct numerical 
simulations. In this situation the invariant is not the magnetic helicity but the anastrophy, \ie the squared magnetic vector potential. 
An inverse cascade was observed when the system is forced at intermediate scale $k_f$ with an associated magnetic energy 
spectrum in $k^{-1}$ compatible with a simple phenomenology \citep{shaikh05}. 
More recently, three-dimensional direct numerical simulations with a mean magnetic field revealed that the propagation of one wave 
packet moving in one direction 
leads to energy transfer towards larger scales \citep{cho11}. This effect interpreted as an inverse cascade (although a constant 
negative flux was not discussed) shows that one dispersive wave packet may produce another wave packet moving in the opposite 
direction whereas the magnetic helicity is well conserved. It is this conservation which is thought to be at the origin of the inverse 
cascade. 

Our paper is devoted to the derivation of new exact solutions for weak KAW/oblique whistler turbulence. \ADD{These solutions imply the 
entanglement of magnetic helicity and energy in the sense that the power law indices of the corresponding spectra are linked 
through a simple relation.}
In section \ref{sect2}, we first develop a discussion about KAW and oblique whistler waves to recall that they are governed by the 
same fluid equations. We conclude that the regime of weak turbulence is the same for both waves. We also derive from compressible 
Hall MHD a compressible version of electron MHD which can be rescaled to give the REMHD. Section \ref{sect3} is the heart of the 
paper: we derive new exact solutions for weak KAW/oblique whistler turbulence. We use the asymptotic 
weak turbulence equations previously derived \citep{galtier03b} and show that the constant magnetic helicity flux spectra are in general 
different from the constant energy flux spectra. These exact solutions allow potentially a magnetic energy spectrum as steep as 
$k_\perp^{-3}$, the limit being fixed by a condition of locality. 
We conclude the paper with a discussion in the last section. It is thought that our results offer a new paradigm to understand solar 
wind turbulence at sub-ion scales where steep magnetic fluctuation power law spectra \ADD{in $f^{\alpha}$ are observed (see figure 
\ref{Fig0}) with a broad range of power law indices such that $\alpha \in [-3.1,-2.5]$ \citep{fouad2013}.}

%%%%%%%%%%%%%%%%%%%%%%%%
\section{KAW and anisotropic/oblique whistler}
\label{sect2}

\ADD{In the introduction, we have explained why Hall MHD and its small scale limit of electron MHD may be relevant to 
describe the solar wind plasma (which is the main domain of application that we consider in this paper and for which some
properties are recalled in the introduction). From this remark, it is believed that it is relevant to make a detailed comparison 
between Hall MHD, electron MHD and REMHD which are often used to investigate solar wind turbulence where in particular
$T_i \sim T_e$.} The goal of this section is thus to recall that the anisotropic version of electron MHD is mathematically 
equivalent to the REMHD and, therefore, the results from weak electron MHD turbulence -- which correspond to a strongly 
anisotropic regime -- are directly applicable to KAW. Additionally, we show that the linearized compressible Hall MHD may exhibit 
a \ADD{simplified} version of the KAW in the low frequencies limit, and that it is possible to derive a compressible version of the 
electron MHD model.

\subsection{Electron MHD}

The incompressible electron MHD equations in presence of an external magnetic field write classically \citep{kingsep}:
\be
\partial_t \bb + d_i b_0 \partial_\parallel (\nabla \times \bb) = - d_i \nabla \times [ (\nabla \times \bb) \times \bb ] \, , 
\label{emhd1}
\ee
\be
\nabla \cdot \bb = 0 \, ,
\label{emhd1b}
\ee
where $\bb$ is a (fluctuating) magnetic field normalized to a velocity ($\bb \to \sqrt{\mu_0 n m_i} \, \bb$, with $m_i$ the ion mass) 
and $\parallel$ is the direction along the external magnetic field $\bb_0 = b_0 {\bf e_\parallel}$. The nonlinear term may take 
another form which is exactly equivalent, namely:
\be
\partial_t \bb + d_i b_0 \partial_\parallel (\nabla \times \bb) = - d_i \nabla \times [ \bb \cdot \nabla \, \bb ] \, . 
\label{emhd2}
\ee
The electron MHD equations can be reduced if we take the anisotropic limit for which $k_\perp \gg k_\parallel$. This limit is 
relevant when an external magnetic field is applied: then, the turbulence may fall in the critical balance regime \citep{cho04} 
or in the weak turbulence regime \citep{galtier03b}. With this limit, we get:
\be
\partial_t \bb + d_i b_0 \partial_\parallel (\nabla_\perp \times \bb) = - d_i \nabla_\perp \times [ \bb_\perp \cdot \nabla_\perp \, \bb ] \, . 
\label{emhd3}
\ee
Expression (\ref{emhd3}) is the anisotropic electron MHD equations. Note that we follow here and below the ordering 
$b_\perp \sim b_\parallel \ll b_0$. The linear solutions of (\ref{emhd3}) are the well-known right-handed anisotropic/oblique 
dispersive whistler waves:
\be
\omega = d_i k_\perp \, b_0 k_\parallel \, . \label{whistler}
\ee

\subsection{Reduced electron MHD}

The REMHD equations have been derived by \cite{sheko09} (see also \cite{boldyrev13}) and below we 
only recall the form of this system when a uniform magnetic field is applied. We obtain ($k_\perp \gg k_\parallel$ is assumed): 
\ba
\partial_t \bb_\perp + {c \over 4 \pi e n_{0e}} b_0 \partial_\parallel (\nabla_\perp \times \bb_\parallel) &=& 
- {c \over 4 \pi e n_{0e}} \nabla_\perp \times \left[ \bb_\perp \cdot \nabla_\perp \, \bb_\parallel \right]  \, , \label{remhd1} \\
\partial_t \bb_\parallel + \lambda {c \over 4 \pi e n_{0e}} b_0 \partial_\parallel (\nabla_\perp \times \bb_\perp) &=& 
- \lambda \, {c \over 4 \pi e n_{0e}} \nabla_\perp \times \left[ \bb_\perp \cdot \nabla_\perp \, \bb_\perp \right]  \, , \label{remhd2}
\ea
where:
\be
\lambda = {\beta_i (1+Z/\tau) \over 2+\beta_i(1+Z/\tau)} \, ,
\ee
$Z=n_e/n_i$ is the charge ratio, $\tau=T_i/T_e$ is the temperature ratio, $\beta_i$ is the ion plasma beta and $n_{0e}$ 
is the constant electron density. It is straightforward to rescale the previous equations by applying the following transform
(we assume $\lambda \neq 0$): 
%(assuming that $\tilde \bb_\perp \sim \bb_\perp$ in order to keep the ordering): 
\be
\bb_\perp \to {\tilde \bb_\perp \over \sqrt{\lambda}} \, , \quad \bb_0 \to {\tilde \bb_0 \over \sqrt{\lambda}} \, ,
\ee
which gives eventually (a simplification is made with the introduction of $d_i$; the gyro-scale $\rho_i$ can also be used with 
the relation $d_i=\rho_i /\sqrt{\beta_i}$): 
\ba
\partial_t \tilde \bb_\perp + d_i \tilde b_0 \partial_\parallel (\nabla_\perp \times \bb_\parallel) 
&=& - d_i \nabla_\perp \times \left[ \tilde \bb_\perp \cdot \nabla_\perp \, \bb_\parallel \right]  \, , 
\label{remhd1b} \\
\partial_t \bb_\parallel + d_i \tilde b_0 \partial_\parallel (\nabla_\perp \times \tilde \bb_\perp) 
&=& - d_i \nabla_\perp \times \left[ \tilde \bb_\perp \cdot \nabla_\perp \, \tilde \bb_\perp \right]  \, . 
\label{remhd2b}
\ea
The addition of both equations leads to (with $\tilde \bb = \tilde \bb_\perp + \bb_\parallel$):
\be
\partial_t \tilde \bb + d_i \tilde b_0 \partial_\parallel (\nabla_\perp \times \tilde \bb)
= - d_i  \nabla_\perp \times \left[ \tilde \bb_\perp \cdot \nabla_\perp \, \tilde \bb \right]  \, , 
\label{remhd3}
\ee
which is mathematically equivalent to the anisotropic version of the electron MHD equations (\ref{emhd3}). 
The linear solutions of (\ref{remhd3}) are the well-known right-handed (anisotropic) dispersive kinetic Alfv\'en waves 
\citep{hasegawa}:
\ba
\omega &=& d_i k_\perp \, \tilde b_0 k_\parallel = \sqrt{\lambda} d_i k_\perp \, b_0 k_\parallel \nonumber \\
&=& \sqrt{\beta_i (1+Z/\tau) \over 2+\beta_i(1+Z/\tau)} \, d_i k_\perp \, b_0 k_\parallel \, .
\label{KAW}
\ea
As expected in the incompressible limit, $\beta_i \to + \infty$, we have $\lambda \to +1$ and we recover exactly expression 
(\ref{whistler}). The same conclusion is reached when $\tau \to 0$.

\subsection{Compressible Hall MHD}
\label{CHMHD}

\ADD{In this section, we shall derive the dispersion relation for compressible Hall MHD and see how a simple version of the KAW 
can be obtained. Compressible Hall MHD equations write:
\ba
{\partial \rho \over \partial t} + \nabla \cdot (\rho \uu) &=& 0 \, , \label{hmhd1} \\
\rho \left({\partial \uu \over \partial t} + \uu \cdot \nabla \uu \right) &=& - \nabla P + 
{1 \over \mu_0}(\nabla \times \bb) \times \bb \, , \label{hmhd2} \\
{\partial \bb \over \partial t} &=& \nabla \times (\uu \times \bb) 
- \nabla \times \left( {(\nabla \times \bb) \times \bb \over \mu_0 n e} \right) \, , \label{hmhd3} \\
\nabla \cdot \bb &=& 0 \, , \label{hmhd4} 
\ea
where $\rho$ is the mass density, $\uu$ the velocity, $P$ the pressure and $n$ the density. With a polytropic closure we have 
$P =  A \rho^\gamma$, where $A$ is a constant of proportionality and $\gamma$ the polytropic index.} 
We focus the analysis on the linear solutions: \ADD{small perturbations (terms with index $1$) are assumed around a constant 
density $\rho_0$, a constant pressure $P_0$ and a constant magnetic field $b_0$.} We have at leading order \ADD{(in Fourier space)}:
\ba
\omega P_1 &=& \rho_0 c_S^2 \, \kk \cdot \uu_1 \, , \\
\omega \uu_1 &=& \kk \left({P_1 \over \rho_0} + \bb_0 \cdot \bb_1 \right) - b_0 \kpa \bb_1 \, , \label{comph1} \\
\omega \bb_1 &=&  - b_0 \kpa \uu_1 + \bb_0 (\kk \cdot \uu_1) + i d_i b_0 \kpa (\kk \times \bb_1) \, , \\
\kk \cdot \bb_1 &=& 0 \, ,
\ea
\ADD{where $c_S=\sqrt{\gamma P_0/\rho_0}$ is the sound speed. Note that the magnetic field has been normalized to a velocity.}
The dispersion relation may be written as: 
\be
{\Omega}^6 - \left(1 + {\beta +1 \over \alpha^2} + K^2 \right) K^2 \alpha^2 {\Omega}^4 
+ \left( 1 + 2 \beta + \beta K^2 \right) K^4 \alpha^2 {\Omega}^2 - \beta K^6 \alpha^4 = 0 \, ,
\label{dishmhdc2}
\ee
with $\Omega \equiv \omega / \omega_{ci}$, $\omega_{ci}=b_0/d_i$, $\alpha \equiv \cos \theta$ (with $\theta$ the angle between 
$\bb_0$ et $\kk$) and $K \equiv k d_i$. This form is interesting for our discussion because in the limit $\alpha \ll 1$ and ${\Omega} \ll 1$, 
it reduces to: 
\be
(\beta +1) {\Omega}^4 - \left( 1 + 2 \beta + \beta K^2 \right) K^2 \alpha^2 {\Omega}^2 + \beta K^4 \alpha^4 = 0 \, . 
\ee
The solutions for $K > 1$ (the term $\propto K^2$ in the second parenthesis increases rapidly with $K$ and does not require
the condition $K \gg 1$ to be dominant) are, ${\Omega} = \sqrt{\beta/(1+\beta)} K^2 \alpha$ and ${\Omega} = \alpha$, which 
corresponds to the following expressions with the previous notations:
\ba
\omega &=& \sqrt{\beta \over 1+\beta} d_i k_\perp b_0 k_\parallel \, , \label{FKAW} \\
\omega &=& \alpha \, \omega_{ci} \, . 
\ea
Whereas the latter expression corresponds to the classical ion cyclotron waves, the former may be interpreted as \ADD{a simplified} 
version of the KAW \ADD{in which the temperatures and the charge ratio do not appear.} 
It corresponds exactly to relation (\ref{KAW}) when $Z/\tau=1$ (which corresponds to $\beta_i=\beta_e$). 
\begin{figure}
\includegraphics{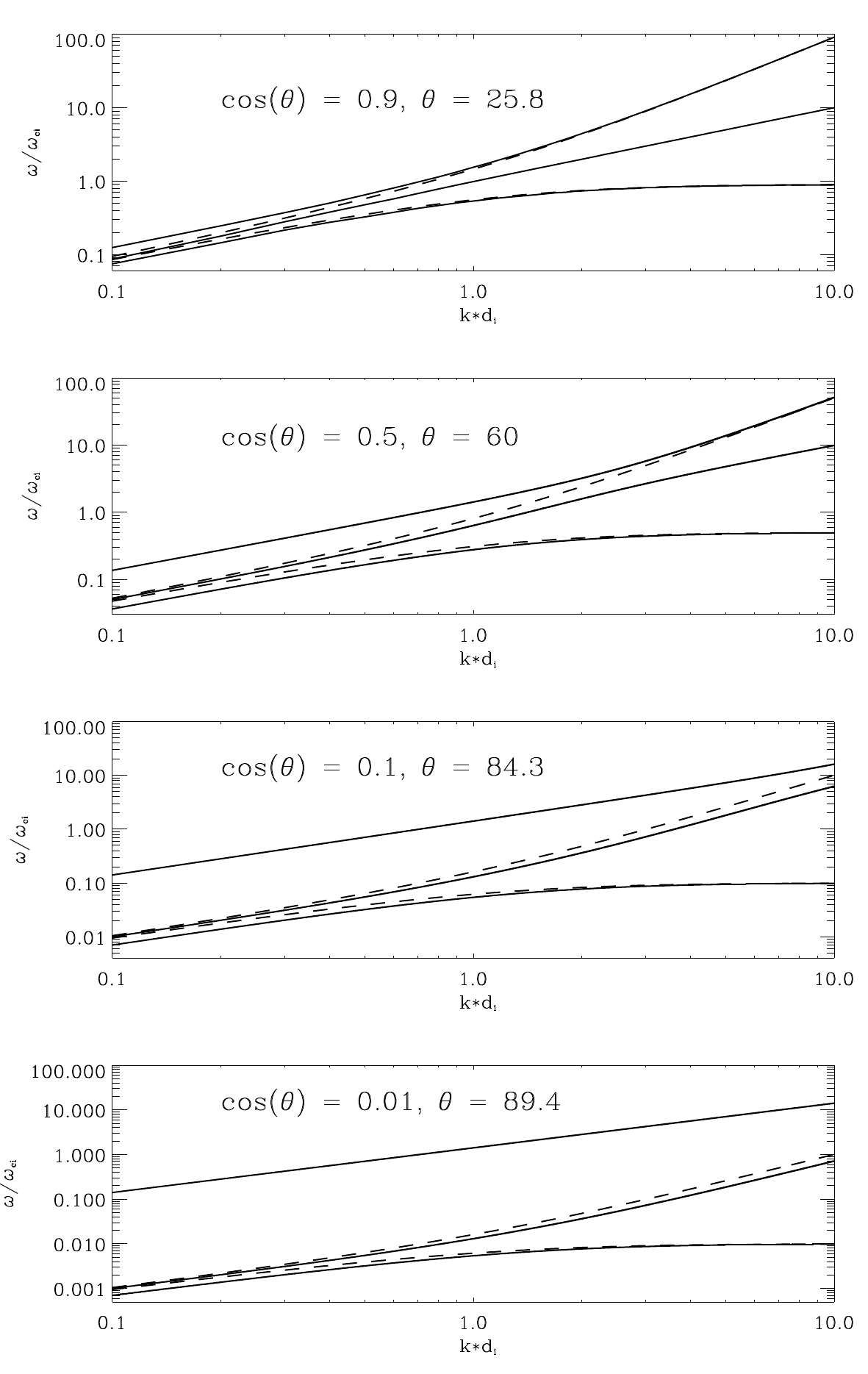} 
\caption{Dispersive branches of compressible Hall MHD (solid) for $\beta=1$ and $\cos \theta = 0.9$, $0.5$, $0.1$ and $0.01$
(from top to bottom respectively). The axes are normalized to $d_i k$ (abscissa) and $\omega / \omega_{ci}$ (ordinate). 
Superimposed are the two branches of incompressible Hall MHD (dash) which defines the right (upper curve) and left (lower curve) 
polarizations. Only a slight difference is seen at large angle between the incompressible and compressible branches of KAW.
Note also the convergence at low wave numbers of the KAW and the incompressible branches to the Alfv\'en branch.}
\label{fig1}
\end{figure}
The validity of this solution may be evaluated by comparing the two last terms of the first parenthesis in expression (\ref{dishmhdc2}). 
For simplicity we assume that $\beta \sim 1$. The KAW solution is found for $K \alpha <1$ whereas we find (with $\Omega \sim 1$) the 
whistler solution for $K \alpha > 1$, which are equivalent to the conditions $k_\parallel d_i  < 1$ and $k_\parallel d_i  > 1$ respectively. We 
may evaluate the critical angle $\theta_c$ which separates both regimes. For that, we consider the electron scale\footnote{\ADD{Although 
the electron inertial term is not included in the Hall MHD equations, we introduce the electron inertial length $d_e$ in the discussion 
as the small-scale limit of the dispersion relation.}}
($k_\perp d_e =1$) and use the relation $1 = k_\parallel d_i = k_\perp d_i / \tan \theta_c$; we obtain $\theta_c = 88.7^o$. In other words, 
for angles larger than $\theta_c$ the right-handed dispersive branch reaches the electron scale as a KAW whereas it is a whistler wave 
for smaller angles. 

The dispersive branches of compressible Hall MHD are shown in figure \ref{fig1} in the particular case of $\beta=1$ (which is of 
interest for solar wind turbulence) and for different 
$\alpha$. We have superimposed the two branches of incompressible Hall MHD to show the right and left polarizations. In the small 
$\alpha$ limit, the kinetic Alfv\'en wave branch appears with a right polarization. We also see that incompressible Hall MHD is 
particularly relevant in the limits $\alpha \sim 1$ and $\alpha \sim 0$.

\subsection{Compressible electron MHD}

We shall derive a compressible version of anisotropic electron MHD (\ref{emhd3}) which can describe the oblique whistler/KAW 
waves (\ref{FKAW}). We start with the compressible Hall MHD equations \ADD{(\ref{hmhd1})--(\ref{hmhd4})} in the isothermal limit 
and assume like for the derivation of REMHD \citep{sheko09} that $\uu \to 0$ but $\nabla \cdot \uu \neq 0$. 
We introduce a uniform magnetic field and find at leading order (after a renormalization) the pressure balance relation:
\be
{P \over \rho_0} + {b_0 b_\parallel} = 0 \, , 
\ee
and:
\be
{\partial \bb \over \partial t} = - \bb_0 \, (\nabla \cdot \uu) - d_i \nabla \times [ \bb \cdot \nabla \, \bb ] \, . 
\ee
The system is closed with the continuity and pressure balance equations; we obtain at leading order (with the isothermal closure 
and the limit $k_\perp \gg k_\parallel$):
\ba
{\partial \bb_\perp \over \partial t} &=& - d_i \nabla_\perp \times [ \bb_\perp \cdot \nabla_\perp \, \bb_\parallel ] \, , \\
{\partial \bb_\parallel \over \partial t} &=& 
- \sqrt{\beta \over 1+ \beta} d_i \nabla_\perp \times [ \bb_\perp \cdot \nabla_\perp \, \bb_\perp ] \, ,
\ea
which are the compressible version of the anisotropic electron MHD equations. As for REMHD, we 
may perform a rescaling transformation. This operation leads to expression (\ref{remhd3}) if we write explicitly the linear term.

\subsection{Conclusion}

\ADD{The KAW cascade can be modeled at several levels of approximation. In particular, we claim that the weak turbulence theory 
previously derived for strongly oblique ($\kpe \gg \kpa$) whistler waves within the electron MHD framework \citep{galtier03b} must be 
interpreted as a weak KAW turbulence theory too. We have seen that (for $\beta=1$) the dispersive branches of incompressible Hall 
MHD follow very well the two lower dispersive branches of compressible Hall MHD when the waves are oblique ($\kpe \gg \kpa$). 
The sonic branch (upper solid curve in the two lowest panels of figure \ref{fig1}) is thought to be less relevant because of its 
damping by kinetic effects \citep{hunana11}. Since oblique waves are the most relevant waves in strongly anisotropic turbulence, 
then we may conclude that in presence of a strong external magnetic field, incompressible Hall MHD offers -- because of its relative 
simplicity -- a more interesting turbulent model than compressible Hall MHD.}

%%%%%%%%%%%%%%%%%%%%%%%%
\section{Exact solutions of weak KAW/whistler turbulence}
\label{sect3}

%%%%%%%%%%%%%%%%%%%%%%%%
\subsection{\ADD{Weak turbulence formalism}}

\ADD{
Weak turbulence is the study of the long time statistical behavior of a sea of weakly nonlinear dispersive waves. It is described 
by wave kinetic equations. In this subsection we present briefly the weak turbulence formalism which leads to these nonlinear 
equations. We shall use the inviscid model equation:
\be
{\partial \bb(\xx,t) \over \partial t} = {\cal L}(\bb) + \epsilon \, {\cal N}(\bb,\bb) \, ,
\label{eq1}
\ee
where $\bb$ is a stationary random vector, ${\cal L}$ is a linear operator which insures that waves 
are solutions of the linear problem, and ${\cal N}$ is a quadratic nonlinear operator (like for electron MHD-type 
fluids). The factor $\epsilon$ is a small parameter ($0 < \epsilon \ll 1$) which will be used for the 
weakly nonlinear expansion.  For electron MHD, the smallness of the nonlinearities 
is the result of the presence of a strong uniform magnetic field ${\bf b_0}$; the operator ${\cal L}$ is 
thus proportional to $b_0$ and $\epsilon \sim b / b_0$ with $b$ the fluctuating magnetic field. 
}

\ADD{
We introduce the three-dimensional direct and inverse Fourier transforms:
\be
\bb(\xx,t) = \int_{\RR^3} \AA(\kk,t) \exp(i \kk \cdot \xx) d\kk \, , 
\ee
\be
\AA(\kk,t) = {1 \over (2\pi)^3} \int_{\RR^3} \bb(\xx,t) \exp(- i \kk \cdot \xx) d\xx \, .
\ee
Therefore, a Fourier transform of equation (\ref{eq1}) gives for the j-component:
\be
\left( {\partial \over \partial t} + i \omega(\kk) \right) A_j (\kk,t) =
\ee
$$ 
\epsilon \int_{\RR^6} \HH_{jmn}^{\kk \pp \qq} A_m(\pp,t) A_n(\qq,t) \delta(\kk-\pp-\qq) d\pp d\qq \, ,
$$
where $\omega(\kk)=\omega_k$ is given by the appropriate dispersion relation (with in general 
$\omega(-\kk) = - \omega(\kk)$) and $\HH$ is a symmetric function in its vector arguments 
which basically depends on the quadratic nonlinear operator ${\cal N}$.
Note the use of the Einstein's notation. We introduce:
\be
\AA(\kk,t) = \aa(\kk,t) e^{-i \omega_k t} \, ,
\ee
and obtain in the interaction representation:
\be
{\partial a_j(\kk) \over \partial t} =
\epsilon \int_{\RR^6} \HH_{jmn}^{\kk \pp \qq} a_m(\pp) a_n(\qq) 
e^{i \Omega_{k,pq}t} \delta_{k,pq} d\pp d\qq \, ,
\label{eq6}
\ee
where the Dirac delta function $\delta_{k,pq}=\delta(\kk-\pp-\qq)$ (then, the wavevectors $\kk$, $\pp$ and $\qq$ form 
a triad) and $\Omega_{k,pq}=\omega_k -\omega_p -\omega_q$;  the time dependence in fields, $\aa$, is omitted for 
simplicity. Relation (\ref{eq6}) is the wave amplitude equation whose dependence in $\epsilon$ means 
that weak nonlinearities will modify only slowly in time the wave amplitude. By nature, the problems
considered here (KAW/whistler waves) involve mainly three-wave interaction processes as it is 
expected by the form of the wave amplitude equation. The exponentially oscillating term is essential for the 
asymptotic closure since we are interested in the long time statistical behavior for which the nonlinear transfer 
time is much greater than the wave period. In such a limit most of the nonlinear terms will be destroyed by 
random phase mixing and only a few of them -- called the resonance terms -- will survive. Before going to the 
statistical formalism, we note the following general properties that will be used:
\ba
\HH_{jmn}^{\kk \pp \qq} &=& \big(\HH_{jmn}^{-\kk -\pp -\qq} \big)^* \, , \\
\HH_{jmn}^{\kk \pp \qq} &{\rm is}& {\rm \, symmetric \, in \,} (\pp, \qq) \, {\rm and \,} (m, n) \, , 
\label{proper} \\
\HH_{jmn}^{0 \pp \qq} &=& 0 \, ,
\ea
where, *, stands for the complex conjugate.
}

\ADD{
We turn now to the statistical description, introduce the ensemble average $\langle ... \rangle$ and 
define the density tensor $q_{jj^\prime}$ for homogeneous turbulence: 
\be
\langle a_j(\kk)a_{j^\prime}(\kk^\prime) \rangle = q_{jj^\prime}(\kk^\prime) \delta(\kk+\kk^\prime) \, . 
\ee
We also assume that on average $\langle \bb(\xx,t) \rangle = 0$ which leads to the relation 
$\HH_{jmn}^{0 \pp \qq}=0$. From the nonlinear equation (\ref{eq6}), we find:
\be
{\partial q_{jj^\prime} \delta(k+k^\prime)\over \partial t} = 
\left\langle a_{j^\prime}(\kk^\prime) {\partial a_j(\kk) \over \partial t} \right\rangle + 
\left\langle a_j(\kk) {\partial a_{j^\prime}(\kk^\prime) \over \partial t} \right\rangle = 
\label{eq8}
\ee
$$
\epsilon \int_{\RR^6} \HH_{jmn}^{\kk \pp \qq} 
\big \langle a_m(\pp) a_n(\qq) a_{j^\prime}(\kk^\prime) \big\rangle 
e^{i \Omega_{k,pq}t} \delta_{k,pq} d\pp d\qq
$$
$$+$$
$$
\epsilon \int_{\RR^6} \HH_{j^{\prime}mn}^{\kk^\prime \pp \qq} 
\big\langle a_m(\pp) a_n(\qq) a_j(\kk) \big\rangle 
e^{i \Omega_{k^\prime,pq}t} \delta_{k^\prime,pq} d\pp d\qq \, .
$$
A hierarchy of equations will clearly appear which gives for the third order moment equation:
\be
{\partial  \big\langle a_j(\kk)a_{j^\prime}(\kk^\prime) a_{j^{\prime\prime}}(\kk^{\prime\prime}) \big\rangle 
\over \partial t} = 
\label{eq123}
\ee
$$
\epsilon \int_{\RR^6} \HH_{jmn}^{\kk \pp \qq}  
\big\langle a_m(\pp) a_n(\qq) a_{j^\prime}(\kk^\prime) a_{j^{\prime\prime}}(\kk^{\prime\prime}) \big\rangle 
e^{i \Omega_{k,pq} t} \delta_{k,pq} d\pp d\qq
$$
$$
+ \, \epsilon \int_{\RR^6} \Big\{ (\kk,j) \leftrightarrow (\kk',j') \Big\} d\pp d\qq
$$
$$
+ \, \epsilon \int_{\RR^6} \Big\{ (\kk",j") \leftrightarrow (\kk',j') \Big\} d\pp d\qq \, ,
$$
where in the right hand side the second line means an interchange in the notations between two pairs with the first 
line as a reference, and the third line means also an interchange in the notations between two pairs with the second 
line as a reference. 
At this stage, we may write the fourth order moment in terms of a sum of the fourth order cumulant 
plus products of second order ones, but a natural closure arises for times asymptotically large 
\citep{nazarenko11}. In this case, several terms do not contribute at large times like, 
in particular, the fourth order cumulant which is not a resonant term. In other words, the 
nonlinear regeneration of third order moments depends essentially on products of second order
moments. The time scale separation imposes a condition of applicability of wave turbulence which 
has to be checked {\it in fine}. After integration in time, we are left with:
\be
\big\langle a_j(\kk)a_{j^\prime}(\kk^\prime) a_{j^{\prime\prime}}(\kk^{\prime\prime}) \big\rangle = 
\ee
$$
\epsilon \int_{\RR^6} \HH_{jmn}^{\kk \pp \qq} 
\Big( \langle a_m(\pp) a_n(\qq) \rangle 
\langle a_{j^\prime}(\kk^\prime) a_{j^{\prime\prime}}(\kk^{\prime\prime}) \rangle
+
\langle a_m(\pp) a_{j^\prime}(\kk^\prime) \rangle
\langle a_n(\qq) a_{j^{\prime\prime}}(\kk^{\prime\prime}) \rangle
$$
$$
+
\langle a_m(\pp) a_{j^{\prime\prime}}(\kk^{\prime\prime}) \rangle
\langle a_n(\qq) a_{j^\prime}(\kk^\prime) \rangle \Big)
\Delta(\Omega_{k,pq}) \delta_{k,pq} d\pp d\qq
$$
$$
+ \, \epsilon \int_{\RR^6} \Big\{ (\kk,j) \leftrightarrow (\kk',j') \Big\} d\pp d\qq
+ \, \epsilon \int_{\RR^6} \Big\{ (\kk",j") \leftrightarrow (\kk',j') \Big\} d\pp d\qq \, ,
$$
where:
\be
\Delta(\Omega_{k,pq}) = \int_0^{t\gg1/\omega} e^{i \Omega_{k,pq}t^\prime} dt^\prime 
=  {e^{i \Omega_{k,pq}t} - 1 \over i \Omega_{k,pq} } \, .
\ee
The same convention as in (\ref{eq123}) is used. After integration in wave vectors $\pp$ and $\qq$ and simplification, 
we get:
\be
\big\langle a_j(\kk)a_{j^\prime}(\kk^\prime) a_{j^{\prime\prime}}(\kk^{\prime\prime}) \big\rangle = 
\ee
$$
\epsilon \Delta(\Omega_{k k^\prime k^{\prime\prime}}) \delta_{kk^\prime k^{\prime\prime}}
$$
$$
\Big( \HH_{jmn}^{\kk -\kk^\prime -\kk^{\prime\prime}} 
q_{mj^\prime}(\kk^\prime) q_{n j^{\prime\prime}} (\kk^{\prime\prime})
+  \HH_{jmn}^{\kk -\kk^{\prime\prime} -\kk^\prime} 
q_{m j^{\prime\prime}} (\kk^{\prime\prime}) q_{n j^\prime} (\kk^\prime)
$$
$$
+ \HH_{j^\prime mn}^{\kk^\prime -\kk -\kk^{\prime\prime}}
q_{mj} (\kk) q_{nj^{\prime\prime}} (\kk^{\prime\prime})
+ \HH_{j^\prime mn}^{\kk^\prime -\kk^{\prime\prime} -\kk} 
q_{mj^{\prime\prime}} (\kk^{\prime\prime}) q_{nj} (\kk)
$$
$$
+ \HH_{j^{\prime\prime}mn}^{\kk^{\prime\prime} -\kk -\kk^\prime}  
q_{mj} (\kk)q_{nj^{\prime}} (\kk^\prime)
+ \HH_{j^{\prime\prime}mn}^{\kk^{\prime\prime} -\kk^\prime -\kk}  
q_{mj^{\prime}} (\kk^\prime) q_{nj} (\kk) \Big) \, .
$$
The symmetries (\ref{proper}) lead to:
\be
\big\langle a_j(\kk)a_{j^\prime}(\kk^\prime) a_{j^{\prime\prime}}(\kk^{\prime\prime}) \big\rangle = 
\ee
$$
2 \epsilon \Delta(\Omega_{k k^\prime k^{\prime\prime}}) \delta_{kk^\prime k^{\prime\prime}}
\Big( \HH_{jmn}^{\kk -\kk^\prime -\kk^{\prime\prime}} 
q_{mj^\prime}(\kk^\prime) q_{n j^{\prime\prime}} (\kk^{\prime\prime})
$$
$$
+ \HH_{j^\prime mn}^{\kk^\prime -\kk -\kk^{\prime\prime}}
q_{mj} (\kk) q_{nj^{\prime\prime}} (\kk^{\prime\prime})
+ \HH_{j^{\prime\prime}mn}^{\kk^{\prime\prime} -\kk -\kk^\prime}  
q_{mj} (\kk)q_{nj^{\prime}} (\kk^\prime) \Big) \, .
$$
The latter expression may be introduced into (\ref{eq8}). We take the long time limit (which introduces 
irreversibility) and find: 
\be
\Delta(x) \to \pi \delta(x) + i {\cal P} (1/x) \, , 
\ee
with ${\cal P}$ the  principal value of the integral. We finally obtain the asymptotically exact wave kinetic 
equations:
\be
{\partial q_{jj^\prime} (\kk) \over \partial t} = 
4 \pi \epsilon^2 \int_{\RR^6} 
\delta_{k,pq} \delta(\Omega_{k,pq}) \HH_{jmn}^{\kk \pp \qq} 
\label{eqfin}
\ee
$$
\Big( \HH_{m r s}^{\pp -\qq -\kk} q_{rn}(\qq) q_{j^\prime s}(\kk)
+ \HH_{nrs}^{\qq -\pp \kk} q_{rm}(\pp) q_{j^\prime s} (\kk) 
+ \HH_{j^\prime rs}^{-\kk -\pp - \qq} q_{rm}(\pp) q_{sn}(\qq) \Big) d\pp d\qq \, .
$$
These general three dimensions wave kinetic equations are valid in principle for any situation where three-wave interaction 
processes are dominant; only the form of $\HH$ has to be adapted to the problem. Equation for the (total) 
energy is obtained by taking the trace of the tensor density, $q_{jj}(\kk)$, whereas other inviscid invariants 
are found by including non diagonal terms. 
}

\subsection{\ADD{Weak turbulence in electron MHD}}

\ADD{As explained above, in electron MHD the natural small parameter is defined from the strong uniform magnetic field such that
$\epsilon \sim b / b_0$.} The preliminary work to such asymptotic developments is the derivation, 
\ADD{from equations (\ref{emhd1})--(\ref{emhd1b}),} of the dynamical equation 
\ADD{(\ref{eq6})} for the wave amplitudes from which we can obtain the resonance conditions. Several properties of weak turbulence 
may be predicted when we study the resonance conditions \citep{Galtier2001}. 
For electron MHD, the nature of the triad interactions has already been investigated \citep{galtier03b,lyutikov} and the analysis shows, 
in general, that the fluid bi-dimensionalises \ADD{with a cascade mostly generated in the direction perpendicular to ${\bf b_0}$.} 
From the wave amplitude equation we may derive the wave kinetic equations \ADD{(\ref{eqfin})} 
governing the long-time behavior of second order 
moments (in our case the magnetic energy and helicity spectra). The achievement of any weak turbulence theory is the derivation of 
such equations with their properties like the exact power law solutions. Contrary to a simple heuristic description, the weak turbulence 
theory offers the possibility to prove rigorously the validity of the power law spectra and to check the locality of the solutions. In addition, 
the sign of the fluxes may be found which gives the direction of the cascade. The latter point is particularly important, first, for the 
comparison with existing data and, second, because it is impossible to predict that from a simple phenomenology.

\subsection{\ADD{Constant magnetic energy flux solutions: previous work}}

The theory of weak (reduced) electron MHD turbulence was derived in \citet{galtier03b} (see also \cite{galtier06} in the context of Hall 
MHD), it is therefore useless to re-derive it. We make the choice to directly recall the wave kinetic equations which describe the time 
evolution of the magnetic energy spectrum:
\ADD{
\be
E(\kk) = {1 \over 2} \langle \bb(\kk) \cdot \bb^*(\kk) \rangle = e^+(\kk) + e^-(\kk) \, , 
\label{defa}
\ee
and magnetic helicity spectrum:
\be
H(\kk) = {1 \over 2} \langle \bb(\kk) \cdot {\bf {\cal A}^*(k)} \rangle = {1 \over k} (e^+(\kk) - e^-(\kk)) \, , 
\label{defb}
\ee
where ${\bf {\cal A}}$ is the vector potential ($\bb = \nabla \times {\bf {\cal A}}$) and $e^s(\kk)$ is the energy 
density tensor introduced by \citet{galtier03b} (equation (40)).}
In the anisotropic limit (which corresponds to the $\kpn \gg \kpa$ limit, and thus to the REMHD case), we have: 
\ba
\partial_t {E_k \brace H_k} &=& {\epsilon^2 \over 16} \, \sum_{s s_p s_q} \int \, {s_p \, \ppn \kpa \ppa \over \qpn} 
\left({s_q \qpn - s_p \ppn \over \kpa} \right)^2 (s \kpn + s_p \ppn + s_q \qpn)^2 \, \sin \theta_q \nonumber \\
\nonumber \\
&&{ s \kpn \left[ E_q ( \ppn E_k - \kpn E_p )/ (\kpn \ppn \qpn) + \, s_q \, H_q \, ( s H_k - s_p H_p ) \right]
\brace
E_q ( s \, H_k - s_p H_p)/\qpn + s_q \, H_q ( \ppn E_k - \kpn E_p )/(\kpn \ppn)} \nonumber \\
\nonumber \\
&&\delta(\kpa + \ppa + \qpa) \, \delta(s\kpn\kpa + s_p\ppn\ppa + s_q\qpn\qpa) \, \diso \, . \label{iso}
\ea
In these equations $E_k \equiv E(\kpn,\kpa)$ and $H_k \equiv H(\kpn,\kpa)$ are respectively the axisymmetric 
bi-dimensional magnetic energy and helicity spectra ($\perp$ and $\parallel$ are respectively the directions perpendicular and 
parallel to $\bf b_0$), $\theta_q$ is the angle between the perpendicular wave vectors $\kkp$ and ${\bf \ppn}$ in the 
triangle made with ($\kkp$, ${\bf \ppn}$, ${\bf \qpn}$) and ($s$, $s_p$, $s_q$) are the directional polarities which 
are equal to $\pm$ (by definition $s \kpa \ge 0$). 
In Eq. (\ref{iso}) the integration over perpendicular wave numbers is such that the triangular relation 
$\kkp + {\bf \ppn} +{\bf \qpn}={\bf 0}$ must be satisfied. 

The solutions of Eq. (\ref{iso}) were previously derived for a turbulence dominated by a forward energy flux \citep{galtier03b}. 
In this case, only the energy equation is useful and the exact finite flux solutions -- obtained by applying a bi-homogeneous 
conformal transformation \citep{ZLF} -- are:
\ba
E_k &\sim& \kpn^{n} \vert \kpa\vert ^{m} \, , \label{KZK1} \\
H_k &\sim& \kpn^{\tilde n} \vert \kpa\vert ^{\tilde m} \, ,
\label{KZK2}
\ea
with $n=-5/2$, $m=-1/2$, $\nnn=-7/2$ and $\mmm=-1/2$. 
Additionally, we can derive the statistically equilibrium solutions for which the energy flux is null; in this case, we have 
$n=1$, $m=0$, $\nnn=0$ and $\mmm=0$.

\subsection{\ADD{Constant magnetic helicity flux solutions: new solutions}}

In a situation where the turbulence is the subject of an inverse magnetic helicity flux it is necessary to consider the second 
equation for the helicity to derive the new exact power law solutions. Also, we implicitly assume that the helicity flux injection 
is made at scale $k_f$ such that the relation $\kpe \gg \kpa$ is satisfied. 
We apply the bi-homogeneous conformal transformation (also called Kuztnesov--Zakharov transform) which consists in doing 
the following manipulation on the wave numbers $\ppn$, $\qpn$, $\ppa$ and $\qpa$:
\be
\begin{array}{lll}
\ppn &\to & \kpn^2 / \ppn \, , \\[.2cm]
\qpn &\to & \kpn \qpn / \ppn \, , \\[.2cm]
\vert \ppa \vert &\to & \kpa^2 / \vert \ppa\vert  \, , \\[.2cm]
\vert \qpa \vert &\to & \vert \kpa\vert \vert  \qpa\vert  / \vert \ppa \vert  \, . 
\end{array}
\label{trans}
\ee
We seek stationary solutions in the power law form (\ref{KZK1})--(\ref{KZK2}) where the parallel components are taken positive. 
After substitution, transformation and simplification, we obtain finally \ADD{(see the derivation in Appendix \ref{appA})}:
\ba
{\partial H_k \over \partial t} \ADD{=} \int && \,
\left[E_0 \left(1 - \left({\ppn \over \kpn}\right)^{n-1} \left\vert{\ppa \over \kpa}\right\vert^{m} \right)
+ H_0 \left(1 - \left({\ppn \over \kpn}\right)^{\nnn} \left\vert{\ppa \over \kpa}\right\vert^{\mmm} \right) \right] \label{KZt} \\
&& \left( 1 - \left({\ppn \over \kpn}\right)^{-n - \nnn - 6} \left\vert{\ppa \over \kpa}\right\vert^{-m - \mmm - 1} \right) \diso \, , 
\nonumber
\ea
where $E_0$ and $H_0$ are some (sophisticated) coefficients which are not relevant to write explicitly. 
The zero helicity flux solutions correspond to the cancellation of both members of the integral in the first line; it gives
$n=1$, $m=0$, $\nnn=0$ and $\mmm=0$. Note that these solutions are exactly the same as those derived 
from the energy equation. The most interesting solutions are, however, those for which the constant helicity flux is finite. 
In this case, we find the relations:
\ba
n+\nnn &=& -6 \, , \label{sol1} \\
m + \mmm &=& -1 \, . \label{sol2}
\ea
\ADD{These solutions show an {\it entanglement} of helicity and energy in the sense that the scaling of one spectrum imposes 
the scaling for the other spectrum.}

\ADD{In this problem, the cascade along the uniform magnetic field is strongly reduced \citep{galtier03b}. 
Simple arguments to explain this property may be found from the resonance condition which can be written as:
\be
\frac{s_p p - s k}{\qpa} = 
\frac{s k - s_q q}{\ppa} =
\frac{s_q q - s_p p}{\kpa} \, .
\label{resonance}
\ee
By considering that the nonlinear transfer is mainly due to local interactions (\ie $k \approx p \approx q$), the resonance 
condition simplifies to: 
\be
{s_p -s \over \qpa} \approx {s_p - s_q \over \kpa} \approx {s - s_q \over \ppa} \, . 
\ee
From the weak turbulence equations, we see that only the interactions between two waves ($\pp$ and $\qq$) with opposite polarities 
($s=s_p=-s_q$ or $s=-s_p=s_q$) will contribute significantly to the nonlinear dynamics. It implies that either $\qpa \approx 0$ or 
$\ppa \approx 0$ which means that only a small transfer is allowed along ${\bf b_0}$. Thus, we may conclude that (i) the local 
nonlinear interactions lead to anisotropic turbulence where the cascade is preferentially generated perpendicularly to ${\bf b_0}$, 
and (ii) the approximation is particularly well verified initially if the turbulence is mainly excited in a limited band of scales since 
then, by nature the nonlinear interactions will be local. 
}
Since the cascade along the uniform magnetic field is strongly reduced, the most important scaling law 
\ADD{in the exact solutions previously derived} is therefore the one for the perpendicular wave numbers. 

It is important to look at the domain of convergence of the integral to check the degree of locality of the power law solutions. 
\ADD{The study of this convergence (see Appendix \ref{appB}) gives} the locality conditions:
\ba
-3 < n + m  < -2 \, , \label{d1} \\
-4 < \nnn + \mmm < -3 \, . \label{d2} 
\ea
We see that with the previous solutions (obtained from the energy or the helicity equations) we are at the border line 
of the domain of convergence. However, we also know that this problem is strongly anisotropic and the inertial range 
in the parallel direction is strongly reduced with a cascade almost only in the perpendicular direction. 
\ADD{Then, the contribution of the power law indices $m$ and $\tilde m$ becomes mainly irrelevant for the 
convergence analysis since it cannot produce any divergence. Note that if we forget this contribution and take\footnote{\ADD{A 
rigorous treatment requires to come back to the weak turbulence equation and introduce for example a delta function 
$\delta(\kpa-k_0)$ in expression (\ref{KZK1}) and (\ref{KZK2}), instead of a power law, in order to model a turbulence without 
inertial zone in the parallel direction.}} simply $m=\mmm=0$,} 
we obtain a classical result of weak turbulence in the sense that the power law indices of the exact solutions 
(\ref{KZK1})--(\ref{KZK2}) fall then at the middle of the domains of locality (\ref{d1})--(\ref{d2}). 
\ADD{Note that a similar situation is found for fast rotating hydrodynamic turbulence (weak inertial wave turbulence regime)}
where direct numerical simulations show an excellent agreement with the weak turbulence predictions (see \cite{galtier03i,galtier14} 
and the references therein). In this case an entanglement \ADD{relation is found (similar to relations (\ref{sol1})--(\ref{sol2}))} 
between the kinetic energy and kinetic helicity. 
\ADD{Note that it is also possible to evaluate the sign of the helicity flux corresponding to the exact power law solutions 
(see Appendix \ref{appC}).}
%which is negative and thus compatible with an inverse cascade.}

%%%%%%%%%%%%%%%%%%%%%%%%
\section{Discussion}
\label{sect4}

The main result of this work is the derivation of the entanglement relations (\ref{sol1})--(\ref{sol2}) for a constant and finite 
magnetic helicity flux. This family of solutions generalizes the spectra derived in the past from the energy equation which 
corresponds to the weak energy cascade of KAW/oblique whistler turbulence. 

\ADD{Our theoretical predictions for weak turbulence show a difference with the result obtained from a two-dimensional direct 
numerical simulation \citep{shaikh05}. The main reason is that only the strong isotropic turbulence regime was investigated 
numerically for which an inverse cascade of anastrophy (the second inviscid invariant of two-dimensional electron MHD) is expected. 
This cascade leads to a magnetic energy spectrum in $k^{-1}$ at large scales compatible with an isotropic phenomenology.}
For the inverse magnetic helicity cascade, the present study may give an interesting limit for the perpendicular scaling assuming that 
the parallel transfer is negligible. Indeed, the convergence condition (\ref{d1}) does not allow an energy spectrum steeper than the 
helicity spectrum with at best a convergence of both power law indices to $n=\nnn=-3$. This limit for the helicity is actually supported 
by a simple anisotropic phenomenology where the wave time writes (see relation (\ref{whistler})):
\be
\tau_{\rm w} \ADD{\sim 1/ \omega} \sim 1 / (\ADD{d_i} b_0 \kpn \kpa) \, .
\ee
The stochastic collisions of KAW/oblique whistler lead to the following estimate for the helicity flux (we mainly consider local interactions 
and use the scaling relation $b \sim k_\perp a$, with $a$ the vector potential \ADD{($\bb = \nabla \times {\bf a}$)}, which also corresponds 
to a maximal helicity state):
\be
\tilde \epsilon \sim {H \over \tau_{\rm eddy}^2/ \tau_{\rm w}} 
\sim {\kpn \kpa H_k \over \tau_{\rm eddy}^2/ \tau_{\rm w}} \, , 
\ee
where $\tau_{\rm eddy} \sim 1/(\ADD{d_i}\kpn^2 b)$
\ADD{is the nonlinear time scale and $H =\int H(\kk) d\kk$ (see relation (\ref{defb})) is the magnetic helicity of the system, \ie a 
spectrum integrated over the three-dimensional Fourier space}; hence, the magnetic helicity spectrum prediction:
\be
H_k \sim \sqrt{{\tilde \epsilon \, b_0 \over \ADD{d_i}}} \, \kpn^{-3} \kpa^{-1/2} \, . \label{moins3}
\ee
Note the limitation of the phenomenology since it is impossible to derive the exact relations (\ref{sol1})--(\ref{sol2}) from expression 
(\ref{moins3}). The exact solutions of weak turbulence are thus highly non-trivial. 

\ADD{Is this weak turbulence regime intermittent or monofractal ? At this level of analysis we cannot answer the question without 
the help of direct numerical simulations. We may predict, however, what would be the scaling laws for higher-order statistics if weak 
turbulence of KAW/whistler is mono fractal. According to the present analysis, the small scales driven by a direct magnetic energy 
cascade is expected to follow the linear relation:}
\be
\zeta_p = {3p \over 4} \, ,
\ee
with by definition $\langle (\bf b (\bf x_\perp + \bf \bell_\perp) - \bf b (\bf x_\perp ))^p \rangle \sim \ell_\perp^{\zeta_p}$, where 
$\bf \bell_\perp$ has to be seen as a vector perpendicular to $\bf b_0$. For the large scales driven by an inverse magnetic helicity 
cascade it is expected to find a solution among a family of linear relations such that, 
\be
{p \over 2} < \zeta_p < p \, . 
\ee
The existence of such double scaling is under numerical investigation and will be presented elsewhere. 

\ADD{The results} presented here may be relevant for solar wind turbulence where the ion and sub-ion scales are now well resolved 
by spacecraft instruments \citep{alexandrova09,Kiyani,Tessein,sahraoui10,bourouaine12,chen13}. 
\ADD{These observations lead to two important questions:}

\ADD{$\bullet$ What is the origin of the scaling laws of the magnetic fluctuation spectra observed in particular after the 
break $f_2$ (see figure \ref{Fig0})?, and}

\ADD{$\bullet$ Why do we observe a wide range of power law indices (between $-2.5$ and $-3.1$) for such spectra?}

\ADD{Our study might give an answer to the second question by assuming that the answer to the first question is {\it turbulence}. 
Indeed, if we assume that the kinetic effects have a negligible contribution on the statistics of the magnetic fluctuations, then the scaling 
laws may be seen as the signature of a turbulence cascade only. Our study suggests that the wide range of values observed for the 
magnetic spectrum power law indices may find its origin in the magnetic helicity and its inverse cascade.}
Since signatures of a non-zero reduced magnetic helicity have been reported at sub-ion scales \citep{howes10}, it would be interesting 
to check if a negative magnetic helicity flux is also present. 
\ADD{Our results is also interesting because for the first time a theory is able to predict rigorously steep power laws for the magnetic 
fluctuation spectrum at sub-ion scale: indeed, previous theories based on the energy cascade were mostly able to propose an index 
of $-7/3$ for strong turbulence or $-2.5$ for weak turbulence. Recently a spectrum close to $-8/3$ has been found numerically by using 
reduced/anisotropic electron MHD \citep{boldyrev12,meyrand13}, and explained differently by invoking the dimensions of 
the dissipative structures (sheets and filaments respectively). Although the origin of this difference is unclear (since basically the 
equations simulated are the same -- see the discussion in section \ref{sect2}) we may think that the strength of the external magnetic 
field has an important role, \eg in destabilizing the current sheets. Note that 
the existence of filaments for this range of scales is also detected in other simulations \citep{martin,karimabadi13,passot14}.}

\ADD{If the explanation of the wide range of power law indices observed in the solar wind comes from the inverse cascade of
magnetic helicity, then a source for the magnetic helicity flux must be found at small scales.} The origin of that helicity injection could 
be related for example to the disruption of small-scales structures at the electron inertial length or electron gyroscale. 
A generalization of the entanglement relation to the critical balance case would be also very welcome. 

\ADD{To conclude the discussion, we think that our theoretical results about the role of the magnetic helicity on the magnetic energy may 
be relevant for MHD turbulence as well where an inverse cascade of magnetic helicity is possible if a magnetic helicity flux is injected. 
In that case, is it possible to find a wide range of power law indices for the magnetic energy spectrum ? 
Fundamental papers on three-dimensional isotropic MHD turbulence \citep{pouquet76} seems to 
indicate that there is a unique scaling in $k^{-2}$ for the magnetic helicity spectrum and that the maximal helicity state is the unique 
solution. Is it really true ? This question could be reinvestigated through direct numerical simulations.}

%%%%%%%%%%%%%%%%%%%%%%%%
R.M. acknowledges the financial support from the French National Research Agency (ANR) contract 10-JCJC-0403. 
%%%%%%%%%%%%%%%%%%%%%%%%

\appendix

%%%%%%%%%%%%%%%%%%%%%%%%
\section{Kolmogorov-Zakharov-Kuznetsov spectra}
\label{appA}

\ADD{
In this appendix, we give the detail of the derivation of the constant helicity flux solutions (\ref{sol1})--(\ref{sol2}). 
We start from the weak turbulence equations (\ref{iso}):
\be
\partial_t {H_k} = {\epsilon^2 \over 16} \, \sum_{s s_p s_q} \int \, {\ppn \kpa s_p \ppa \over \qpn} 
\left({s_q \qpn - s_p \ppn \over \kpa} \right)^2 (s \kpn + s_p \ppn + s_q \qpn)^2 \, \sin \theta_q \label{isoAA}
\ee
$$
\left[ {E_q \over \qpn} ( s \, H_k - s_p H_p) + {s_q \, H_q \over \kpn \ppn} ( \ppn E_k - \kpn E_p ) \right]
$$
$$
\delta(\kpa - \ppa - \qpa) \, \delta(s \kpn \kpa - s_p \ppn \ppa - s_q \qpn \qpa) \, \diso \, . 
$$
Note in passing that from the definitions (\ref{defa})--(\ref{defb}), we have the relation:
\be
e^\pm(\kk) = {1 \over 2} \left(E(\kk) \pm k H(\kk) \right) \, ,
\ee
for the energy density tensor which is a positive definite quantity; then we obtain the Schwarz inequality 
$k H(\kk) \le E(\kk)$. We define the spectra:
\ba
E(\kpn,\kpa) &=& C_E \kpn^n \vert \kpa \vert^m \, , \\
H(\kpn,\kpa) &=& C_H \kpn^{\tilde n} \vert \kpa \vert^{\tilde m} \, ,
\ea
where $C_E$ and $C_H$ are some constants and we shall look for exact power law solutions of the weak turbulence equations. 
Note that the solutions found with the Kuznetsov-Zakharov transform (see below) are not necessary the unique solutions to this 
problem in the sense that the uniqueness of these solutions is not proved \citep{lvov04,nazarenko11}. 
We introduce the previous expressions into the weak turbulence equations and obtain after simple manipulations (\eg identity 
relation for a triangle is used):
\be
\partial_t {H(\kpn,\kpa >0) } =
\ee
$$
{\epsilon^2 \over 16} \, \sum_{s s_p s_q} \int \, 
\left({\sin \theta_k \over \kpn}\right) \ppn \vert \kpa \vert \vert \ppa \vert 
\left({s_q \qpn - s_p \ppn \over \kpa} \right)^2 (s \kpn + s_p \ppn + s_q \qpn)^2 
$$
$$
C_E C_H \left[ \qpn^{n-1} \vert \qpa \vert^m
( s \kpn^{\tilde n} \vert \kpa \vert^{\tilde m} - s_p \ppn^{\tilde n} \vert \ppa \vert^{\tilde m}) 
+ s_q \qpn^{\tilde n} \vert \qpa \vert^{\tilde m}
( \ppn^{n-1} \vert \ppa \vert^m - \kpn^{n-1} \vert \kpa \vert^m ) \right]
$$
$$
\delta(\kpa - \ppa - \qpa) \, \delta(s \kpn \kpa - s_p \ppn \ppa - s_q \qpn \qpa) \, \diso \, . \label{asplit}
$$
Then, we split the integral into two identical integrals and apply the Kuznetsov-Zakharov transform on one of them. We obtain:
\be
\partial_t {H_k} = {\epsilon^2 \over 32} \, \sum_{s s_p s_q} \int \, 
\left({\sin \theta_k \over \kpn}\right) \ppn \vert \kpa \vert \vert \ppa \vert 
\left({s_q \qpn - s_p \ppn \over \kpa} \right)^2 (s \kpn + s_p \ppn + s_q \qpn)^2 \label{split}
\ee
$$
C_E C_H \left[ \qpn^{n-1} \vert \qpa \vert^m
( s \kpn^{\tilde n} \vert \kpa \vert^{\tilde m} - s_p \ppn^{\tilde n} \vert \ppa \vert^{\tilde m}) 
+ s_q \qpn^{\tilde n} \vert \qpa \vert^{\tilde m}
( \ppn^{n-1} \vert \ppa \vert^m - \kpn^{n-1} \vert \kpa \vert^m ) \right]
$$
$$
\delta(\kpa - \ppa - \qpa) \, \delta(s \kpn \kpa - s_p \ppn \ppa - s_q \qpn \qpa) \, \diso \, 
$$
$$
+ {\epsilon^2 \over 32} \, \sum_{s s_p s_q} \int \, 
\left({\sin \theta_k \over \kpn} {\ppn \over \kpn} \right) {\kpn^2 \over \ppn} \vert \kpa \vert {\vert \kpa \vert^2 \over \vert \ppa \vert}
\left[\left({s_q \qpn - s_p \kpn \over \ppa} \right) {\kpn \over \ppn} {\vert \ppa \vert \over \vert \kpa \vert} \right]^2 
$$
$$
\left( (s_p \kpn + s \ppn + s_q \qpn) {\kpn \over \ppn}\right)^2 C_E C_H 
$$
$$
[ \kpn^{n-1} \qpn^{n-1} \ppn^{-n+1} \vert \kpa \vert^m \vert \qpa \vert^m \vert \ppa \vert^{-m}
( s \kpn^{\tilde n} \vert \kpa \vert^{\tilde m} - s_p \kpn^{2\tilde n} \ppn^{-\tilde n} \vert \kpa \vert^{2\tilde m} \vert \ppa \vert^{-\tilde m}) 
$$
$$
+ s_q \kpn^{\tilde n} \qpn^{\tilde n} \ppn^{-\tilde n} \vert \kpa \vert^{\tilde m} \vert \qpa \vert^{\tilde m} \vert \ppa \vert^{-\tilde m}
( \kpn^{2n-2} \ppn^{-n+1} \vert \kpa \vert^{2m} \vert \ppa \vert^{-m} - \kpn^{n-1} \vert \kpa \vert^m ) ]
$$
$$
\delta(\kpa + \qpa - \ppa) {\vert \ppa \vert \over \vert \kpa \vert} \, 
\delta(s_p \kpn \kpa + s_q \qpn \qpa - s \ppn \ppa) {\ppn \vert \ppa \vert \over \kpn \vert \kpa \vert}
\left({\kpn \vert \kpa \vert \over \ppn \vert \ppa \vert}\right)^3 \, \diso \, . 
$$
In the second integral, we exchange the dummy variables $s$ and $s_p$ and use relation (\ref{resonance}) 
in the anisotropic limit ($\kpn \gg \kpa$); we obtain:
\be
\partial_t {H_k} = {\epsilon^2 \over 32} \, \sum_{s s_p s_q} \int \, 
\left({\sin \theta_k \over \kpn}\right) \ppn \vert \kpa \vert \vert \ppa \vert 
\left({s_q \qpn - s_p \ppn \over \kpa} \right)^2 (s \kpn + s_p \ppn + s_q \qpn)^2 \label{split2}
\ee
$$
C_E C_H \left[ \qpn^{n-1} \vert \qpa \vert^m
( s \kpn^{\tilde n} \vert \kpa \vert^{\tilde m} - s_p \ppn^{\tilde n} \vert \ppa \vert^{\tilde m}) 
+ s_q \qpn^{\tilde n} \vert \qpa \vert^{\tilde m}
( \ppn^{n-1} \vert \ppa \vert^m - \kpn^{n-1} \vert \kpa \vert^m ) \right]
$$
$$
\delta(\kpa - \ppa - \qpa) \, \delta(s \kpn \kpa - s_p \ppn \ppa - s_q \qpn \qpa) \, \diso \, 
$$
$$
+ {\epsilon^2 \over 32} \, \sum_{s s_p s_q} \int \, 
\left({\sin \theta_k \over \kpn}\right) \ppn \vert \kpa \vert \vert \ppa \vert 
\left({s_q \qpn - s_p \ppn \over \kpa} \right)^2 (s \kpn + s_p \ppn + s_q \qpn)^2 \left({\kpn \over \ppn}\right)^5
$$
$$
C_E C_H 
[ \kpn^{n-1} \qpn^{n-1} \ppn^{-n+1} \vert \kpa \vert^m \vert \qpa \vert^m \vert \ppa \vert^{-m}
( s_p \kpn^{\tilde n} \vert \kpa \vert^{\tilde m} - s \kpn^{2\tilde n} \ppn^{-\tilde n} \vert \kpa \vert^{2\tilde m} \vert \ppa \vert^{-\tilde m}) 
$$
$$
+ s_q \kpn^{\tilde n} \qpn^{\tilde n} \ppn^{-\tilde n} \vert \kpa \vert^{\tilde m} \vert \qpa \vert^{\tilde m} \vert \ppa \vert^{-\tilde m}
( \kpn^{2n-2} \ppn^{-n+1} \vert \kpa \vert^{2m} \vert \ppa \vert^{-m} - \kpn^{n-1} \vert \kpa \vert^m ) ]
$$
$$
\delta(\kpa + \qpa - \ppa) \, 
\delta(s \kpn \kpa + s_q \qpn \qpa - s_p \ppn \ppa) \left({\kpn \over \ppn}\right)^2
{\vert \kpa \vert \over \vert \ppa \vert} \, \diso \, . 
$$
After some other manipulations, we find eventually:
\be
\partial_t {H_k} = {\epsilon^2 \over 32} \, \sum_{s s_p s_q} \int 
\left({\sin \theta_k \over \kpn}\right) \ppn \vert \ppa \vert 
\left({s_q \qpn - s_p \ppn \over \kpa} \right)^2 (s \kpn + s_p \ppn + s_q \qpn)^2 \label{split3}
\ee
$$
\left[\left({\qpn \over \kpn}\right)^{n-1} \left\vert {\qpa \over \kpa}\right\vert^m 
\left( s - s_p \left({\ppn \over \kpn}\right)^{\tilde n} \left\vert{\ppa \over \kpa}\right\vert^{\tilde m} \right) 
- s_q \left({\qpn \over \kpn}\right)^{\tilde n} \left\vert{\qpa \over \kpa}\right\vert^{\tilde m} 
\left(1 - \left({\ppn \over \kpn}\right)^{n-1} \left\vert{\ppa \over \kpa}\right\vert^m\right) 
\right]
$$
$$
C_E C_H \kpn^{n+ \tilde n -1} \vert \kpa \vert^{m + \tilde m+1} 
\left(1-\left({\ppn \over \kpn}\right)^{-n-\tilde n -6} \left\vert{\ppa \over \kpa }\right\vert^{-m -\tilde m -1} \right)
$$
$$
\delta(\kpa - \ppa - \qpa) \, \delta(s \kpn \kpa - s_p \ppn \ppa - s_q \qpn \qpa) \, \diso \, .
$$
Then, exact power law solutions may be derived easily: it corresponds to the cancellation of the integrand (\ie stationary solutions). 
The most general solutions (Kolmogorov-Zakharov spectra) are obtained by taking:
\ba
n+\nnn &=& -6 \, , \\
m + \mmm &=& -1 \, . 
\ea
}

%%%%%%%%%%%%%%%%%%%%%%%%
\section{Locality of the interactions}
\label{appB}

\ADD{
This appendix is devoted to the locality of the solutions derived in the previous appendix. 
It is basically a convergence analyzis. We start from the weak turbulence equations (\ref{iso}):
\be
\partial_t {H_k} = {\epsilon^2 \over 16} \, \sum_{s s_p s_q} \int \, {\ppn \kpa s_p \ppa \over \qpn} 
\left({s_q \qpn - s_p \ppn \over \kpa} \right)^2 (s \kpn + s_p \ppn + s_q \qpn)^2 \, \sin \theta_q \label{wte1}
\ee
$$
\left[ {E_q \over \qpn} ( s \, H_k - s_p H_p) + {s_q \, H_q \over \kpn \ppn} ( \ppn E_k - \kpn E_p ) \right]
$$
$$
\delta(\kpa - \ppa - \qpa) \, \delta(s \kpn \kpa - s_p \ppn \ppa - s_q \qpn \qpa) \, \diso \, ,
$$
in which we introduce the power law spectra:
\ba
E(\kpn,\kpa) &=& C_E \kpn^n \vert \kpa \vert^m \, , \\
H(\kpn,\kpa) &=& C_H \kpn^{\tilde n} \vert \kpa \vert^{\tilde m} \, .
\ea
We obtain:
\be
\partial_t {H_k} = {\epsilon^2 \over 16} \, \sum_{s s_p s_q} \int \, {\tppn \over \tqpn} s_p \tppa 
\left({s_q \tqpn - s_p \tppn} \right)^2 (s + s_p \tppn + s_q \tqpn)^2 \, \sin \theta_q 
\ee
$$
\kpn^{n+\tilde n +4} \kpa^{m + \tilde m} C_E C_H 
\left[ \tqpn^{n-1} \tqpa^m (s - s_p \tppn^{\tilde n} \tppa^{\tilde m}) 
+ s_q \tqpn^{\tilde n} \tqpa^{\tilde m} (1- \tppn^{n-1} \tppa^m) \right]
$$
$$
\delta(1 - \tppa - \tqpa) \, \delta(s - s_p \tppn \tppa - s_q \tqpn \tqpa) \, \disot \, ,
$$
where $\tppn \equiv \ppn / \kpn$, $\tqpn \equiv \qpn / \kpn$, $\tppa \equiv \vert \ppa \vert / \vert \kpa \vert $ and 
$\tqpa \equiv \vert  \qpa \vert  / \vert \kpa \vert $. 
For convenience, we shall use the following form:
\be
\partial_t {H_k} = {\epsilon^2 \over 16} \kpn^{n+\tilde n +4} \kpa^{m + \tilde m} C_E C_H 
\sum_{s s_p s_q} \int \sin \theta_q {\tppn \over \tqpn} s_p \tppa 
\left({s_q \tqpn - s_p \tppn} \right)^2 
\ee
$$
(s + s_p \tppn + s_q \tqpn)^2 \left[ \tqpn^{n-1} \tqpa^m (s - s_p \tppn^{\tilde n} \tppa^{\tilde m}) 
+ s_q \tqpn^{\tilde n} \tqpa^{\tilde m} (1- \tppn^{n-1} \tppa^m) \right]
$$
$$
\delta(1 - \tppa - \tqpa) \, \delta(s - s_p \tppn \tppa - s_q \tqpn \tqpa) \, \disot \, .
$$
Since the weak turbulence equation (\ref{wte1}) is only valid in the limit $\kpn \gg \kpa$, we introduce the parameterized variables 
$\tppa = \epsilon_p \tppn$ and $\tqpa = \epsilon_q \tqpn$ for the calculation, where $\epsilon_p \ll 1$ and $\epsilon_q \ll 1$. 
Three non local limits will be analyzed (see figure \ref{Fig2}). 
\begin{figure}
\centerline{\includegraphics[height=7cm]{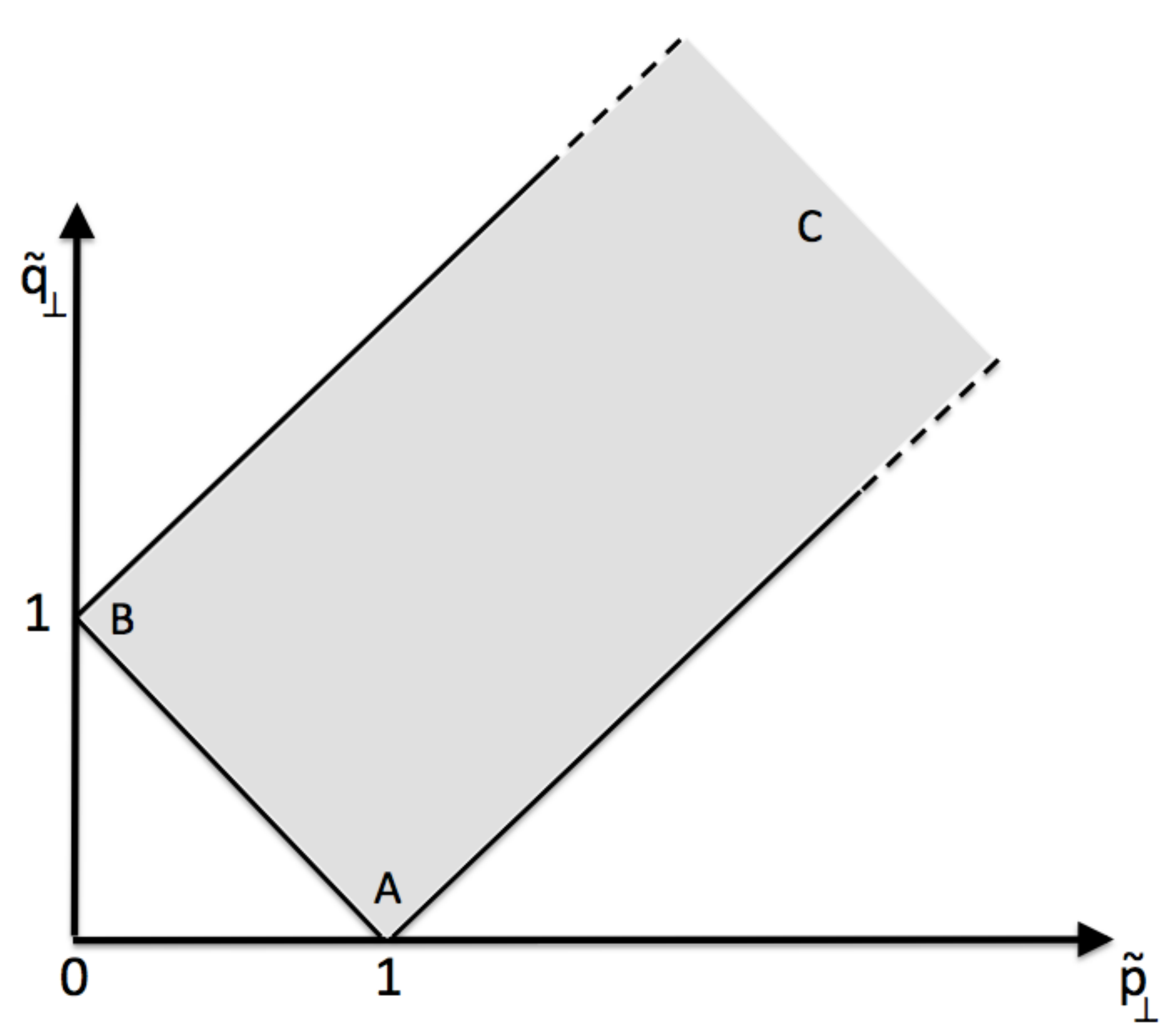}}
\caption{\ADD{The weak turbulence equations are integrated over a domain such that $\kk = \pp + \qq$. The (gray) band corresponds 
to this domain for the normalized perpendicular wave vectors. Points A, B, and C (at infinity) are the regions of non locality.}}
\label{Fig2}
\end{figure}
}

\medskip

\ADD{
$\bullet$ Case A:
\ba
\tppn &=& 1 + r \cos \alpha \, , \\
\tqpn &=& r \sin \alpha \, , 
\ea
where $\alpha \in [\pi/4, 3\pi/4]$ and $r \ll 1$. For this region, we find the convergence condition:
\ba
n + m &>& -3 \, , \\
\tilde n + \tilde m &>& -4 \, .
\ea
}

\ADD{
$\bullet$ Case B:
\ba
\tppn &=& r \cos \alpha \, , \\
\tqpn &=& 1+ r \sin \alpha \, , 
\ea
where $\alpha \in [-\pi/4, \pi/4]$ and $r \ll 1$. For this region, we find the convergence condition:
\ba
n + m &>& -4 \, , \\
\tilde n + \tilde m &>& -5 \, .
\ea
}

\ADD{
$\bullet$ Case C:
\ba
\tppn &=& {\tau_2 - \tau_1 \over 2} \, , \\
\tqpn &=& {\tau_1 + \tau_2 \over 2} \, , 
\ea
where $-1<\tau_1<+1$ and $+1<\tau_2< + \infty$. For this region, we find the convergence condition:
\ba
n + m &<& -2 \, , \\
\tilde n + \tilde m &<& -3 \, . 
\ea
The locality analysis for the parallel wave numbers does not lead to any new constrains.
Thus, the convergence condition corresponds to the inequalities:
\ba
-3 < n &+& m < -2 \, , \\
-4 < \tilde n &+& \tilde m < -3 \, .
\ea
}

%%%%%%%%%%%%%%%%%%%%%%%%
\section{Sign of magnetic helicity flux}\label{appC}

\ADD{
The sign of the helicity flux may be investigated from the weak turbulence equation (\ref{split3}):
\be
\partial_t {H_k} = {\epsilon^2 \over 32} \, \sum_{s s_p s_q} \int 
\left({\sin \theta_k \over \kpn}\right) \ppn \vert \ppa \vert 
\left({s_q \qpn - s_p \ppn \over \kpa} \right)^2 (s \kpn + s_p \ppn + s_q \qpn)^2 \label{split4}
\ee
$$
\left[\left({\qpn \over \kpn}\right)^{n-1} \left\vert {\qpa \over \kpa}\right\vert^m 
\left( s - s_p \left({\ppn \over \kpn}\right)^{\tilde n} \left\vert{\ppa \over \kpa}\right\vert^{\tilde m} \right) 
- s_q \left({\qpn \over \kpn}\right)^{\tilde n} \left\vert{\qpa \over \kpa}\right\vert^{\tilde m} 
\left(1 - \left({\ppn \over \kpn}\right)^{n-1} \left\vert{\ppa \over \kpa}\right\vert^m\right) 
\right]
$$
$$
C_E C_H \kpn^{n+ \tilde n -1} \vert \kpa \vert^{m + \tilde m+1} 
\left(1-\left({\ppn \over \kpn}\right)^{-n-\tilde n -6} \left\vert{\ppa \over \kpa }\right\vert^{-m -\tilde m -1} \right)
$$
$$
\delta(\kpa - \ppa - \qpa) \, \delta(s \kpn \kpa - s_p \ppn \ppa - s_q \qpn \qpa) \, \diso \, .
$$
Additionally, we have \citep{ZLF}:
\be 
\partial_t {H_k} = - \nabla \cdot {\bf P} = -{1 \over \kpe} {\partial (\kpe P_\perp) \over \partial \kpe } 
- {\partial P_\parallel \over \partial \kpa} \, , \label{flux1}
\ee
where ${\bf P}$ is the helicity flux vector , $P_\perp$ and $P_\parallel$ the perpendicular and parallel components 
of this flux vector (axisymmetric turbulence is assumed) respectively. We introduce the notations: 
$\tppn \equiv \ppn / \kpn$, $\tqpn \equiv \qpn / \kpn$, $\tppa \equiv \vert \ppa \vert / \vert \kpa \vert $ and 
$\tqpa \equiv \vert  \qpa \vert  / \vert \kpa \vert $ and obtain:
\be
\partial_t {H_k} = \epsilon^2 C_E C_H \kpe^{n+\tilde n +4} \vert \kpa \vert^{m + \tilde m} I(m,n,\tilde m, \tilde n) \, , 
\label{split5}
\ee
with:
\be
I (m,n,\tilde m, \tilde n) = \label{split6}
\ee
$$
{1 \over 32} \, \sum_{s s_p s_q} \int 
\sqrt{1 - \left({\tppn^2 + \tqpn^2-1 \over 2 \tppn \tqpn}\right)^2}
\, \tppn \tppa \left(s_q \tqpn - s_p \tppn \right)^2 (s + s_p \tppn + s_q \tqpn)^2 
$$
$$
\left[ \tqpn^{n-1} \tqpa^m \left( s - s_p \tppn^{\tilde n} \tppa^{\tilde m} \right) 
- s_q \tqpn^{\tilde n} \qpa^{\tilde m} \left(1 - \tppn^{n-1} \tppa^m\right) \right]
\left(1-\tppn^{-n-\tilde n -6} \tppa^{-m -\tilde m -1} \right)
$$
$$
\delta(1 - \tppa - \tqpa) \, \delta(s - s_p \tppn \tppa - s_q \tqpn \tqpa) \, \disot \, .
$$
From the flux equation (\ref{flux1}), we may have at constant $\kpa$:
\be
{\partial (\kpe P_\perp) \over \partial \kpe } 
= - \epsilon^2 C_E C_H \kpe^{n+\tilde n +5} \vert \kpa \vert^{m + \tilde m} I(m,n,\tilde m, \tilde n) \, .
\ee
After an integration, we have the general relation:
\be
P_\perp = - \epsilon^2 C_E C_H \kpe^{n+\tilde n +5} \vert \kpa \vert^{m + \tilde m} {I(m,n,\tilde m, \tilde n) \over n+\tilde n +6} \, .
\ee
The constant flux solution that we look for corresponds precisely to the cancellations of the denominator and the numerator $I$. 
This indeterminacy can be evaluated using L'Hospital's rule; we find for this solution:
\be
P_\perp = - \epsilon^2 {C_E C_H \over \kpe \vert \kpa \vert} A \, ,
\label{rA}
\ee
with:
\be
A= \left({\partial I(m,n,\tilde m, \tilde n) \over \partial (n+\tilde n +6)}\right)_{n+\tilde n = -6, \, m+\tilde m = -1} = 
\ee
\be
{1 \over 32} \, \sum_{s s_p s_q} \int 
\sqrt{1 - \left({\tppn^2 + \tqpn^2-1 \over 2 \tppn \tqpn}\right)^2}
\, \tppn \tppa \left(s_q \tqpn - s_p \tppn \right)^2 (s + s_p \tppn + s_q \tqpn)^2 
\ee
$$
\ln (\tppn)
\left[ \tqpn^{n-1} \tqpa^m \left( s - s_p \tppn^{-n-6} \tppa^{-m-1} \right) 
- s_q \tqpn^{-n-6} \qpa^{-m-1} \left(1 - \tppn^{n-1} \tppa^m\right) \right]
$$
$$
\delta(1 - \tppa - \tqpa) \, \delta(s - s_p \tppn \tppa - s_q \tqpn \tqpa) \, \disot \, .
$$
%%%%%%%%%%%
In a similar way from the flux equation (\ref{flux1}), we may have at constant $\kpe$:
\be
{\partial P_\parallel \over \partial \kpa} = 
- \epsilon^2 C_E C_H \kpe^{n+\tilde n +4} \vert \kpa \vert^{m + \tilde m} I(m,n,\tilde m, \tilde n) \, .
\ee
After an integration, we find the general relation:
\be
P_\parallel = - \epsilon^2 C_E C_H \kpe^{n+\tilde n +4} \vert \kpa \vert^{m + \tilde m +1} {I(m,n,\tilde m, \tilde n) \over m + \tilde m +1} \, . 
\ee
As above, the constant flux solution corresponds to the cancellations of the denominator and $I$. 
Thanks to L'Hospital's rule, we find:
\be
P_\parallel = 
- \epsilon^2 {C_E C_H \over \kpe^2} B \, ,
\label{rB}
\ee
with (we introduce the solutions (\ref{sol1})--(\ref{sol2})):
\be
B= \left({\partial I(m,n,\tilde m, \tilde n) \over \partial(m + \tilde m +1)}\right)_{n+\tilde n = -6, \, m+\tilde m = -1} = 
\ee
\be
{1 \over 32} \, \sum_{s s_p s_q} \int 
\sqrt{1 - \left({\tppn^2 + \tqpn^2-1 \over 2 \tppn \tqpn}\right)^2}
\, \tppn \tppa \left(s_q \tqpn - s_p \tppn \right)^2 (s + s_p \tppn + s_q \tqpn)^2 
\ee
$$
\ln (\tppa) 
\left[ \tqpn^{n-1} \tqpa^m \left( s - s_p \tppn^{-n-6} \tppa^{-m-1} \right) 
- s_q \tqpn^{-n-6} \qpa^{-m-1} \left(1 - \tppn^{n-1} \tppa^m\right) \right]
$$
$$
\delta(1 - \tppa - \tqpa) \, \delta(s - s_p \tppn \tppa - s_q \tqpn \tqpa) \, \disot \, .
$$
The combination of relations (\ref{rA}) and (\ref{rB}) gives in particular the flux ratio:
\be
{P_\parallel \over P_\perp} = {\kpa \over \kpe} {B \over A} \, ,
\ee
which is small if $A$ and $B$ are of the same order. 
We see that the signs of the fluxes $P_\perp$ and $P_\parallel$ will be given by the signs of the constants $A$ and $B$
respectively (the constants $C_E$ and $C_H$ are taken positive). 
}

\bibliographystyle{jpp}
\bibliography{jpp-galtier}

\begin{thebibliography}{76}
\expandafter\ifx\csname natexlab\endcsname\relax\def\natexlab#1{#1}\fi

\bibitem[{Alexandrova} {\em et~al.\/}(2009){Alexandrova}, {Saur}, {Lacombe},
  {Mangeney}, {Mitchell}, {Schwartz} \& {Robert}]{alexandrova09}
{\sc {Alexandrova}, O., {Saur}, J., {Lacombe}, C., {Mangeney}, A., {Mitchell},
  J., {Schwartz}, S.~J. \& {Robert}, P.} 2009 {Universality of Solar-Wind
  Turbulent Spectrum from MHD to Electron Scales}. {\em Phys. Rev. Lett.\/}
  {\bf 103}~(16), 165003.

\bibitem[{Banerjee} {\em et~al.\/}(2013){Banerjee}, {Ray}, {Sahoo} \&
  {Pandit}]{pandit}
{\sc {Banerjee}, D., {Ray}, S.~S., {Sahoo}, G. \& {Pandit}, R.} 2013
  {Multiscaling in Hall-Magnetohydrodynamic Turbulence: Insights from a Shell
  Model}. {\em Phys. Rev. Lett.\/} {\bf 111}~(17), 174501.

\bibitem[{Bhattacharjee}(2004)]{Bhatt04}
{\sc {Bhattacharjee}, A.} 2004 {Impulsive Magnetic Reconnection in the Earth's
  Magnetotail and the Solar Corona}. {\em Annu. Rev. Astron. Astrophys.\/} {\bf
  42}, 365--384.

\bibitem[{Biskamp}(2000)]{biskamp00}
{\sc {Biskamp}, D.} 2000 {\em Magnetic Reconnection in Plasmas\/}. Cambridge
  Univ. Press, Cambridge.

\bibitem[{Biskamp} {\em et~al.\/}(1996){Biskamp}, {Schwarz} \&
  {Drake}]{biskamp96}
{\sc {Biskamp}, D., {Schwarz}, E. \& {Drake}, J.~F.} 1996 {Two-Dimensional
  Electron Magnetohydrodynamic Turbulence}. {\em Phys. Rev. Lett.\/} {\bf 76},
  1264--1267.

\bibitem[{Boldyrev} {\em et~al.\/}(2013){Boldyrev}, {Horaites}, {Xia} \&
  {Perez}]{boldyrev13}
{\sc {Boldyrev}, S., {Horaites}, K., {Xia}, Q. \& {Perez}, J.~C.} 2013 {Toward
  a Theory of Astrophysical Plasma Turbulence at Subproton Scales}. {\em
  Astrophys. J.\/} {\bf 777}, 41.

\bibitem[{Boldyrev} \& {Perez}(2012)]{boldyrev12}
{\sc {Boldyrev}, S. \& {Perez}, J.~C.} 2012 {Spectrum of Kinetic-Alfv{\'e}n
  Turbulence}. {\em Astrophys. J. Lett.\/} {\bf 758}, L44.

\bibitem[{Bourouaine} {\em et~al.\/}(2012){Bourouaine}, {Alexandrova}, {Marsch}
  \& {Maksimovic}]{bourouaine12}
{\sc {Bourouaine}, S., {Alexandrova}, O., {Marsch}, E. \& {Maksimovic}, M.}
  2012 {On Spectral Breaks in the Power Spectra of Magnetic Fluctuations in
  Fast Solar Wind between 0.3 and 0.9 AU}. {\em Astrophys. J.\/} {\bf 749},
  102.

\bibitem[{Bulanov} {\em et~al.\/}(1992){Bulanov}, {Pegoraro} \&
  {Sakharov}]{bulanov}
{\sc {Bulanov}, S.~V., {Pegoraro}, F. \& {Sakharov}, A.~S.} 1992 {Magnetic
  reconnection in electron magnetohydrodynamics}. {\em Phys. Fluids B\/} {\bf
  4}, 2499--2508.

\bibitem[{Cai} {\em et~al.\/}(2011){Cai}, {Zhu}, {Chen}, {Wu}, {He} \&
  {Mima}]{cai}
{\sc {Cai}, H.-B., {Zhu}, S.-P., {Chen}, M., {Wu}, S.-Z., {He}, X.~T. \&
  {Mima}, K.} 2011 {Magnetic-field generation and electron-collimation analysis
  for propagating fast electron beams in overdense plasmas}. {\em Phys. Rev.
  E\/} {\bf 83}~(3), 036408.

\bibitem[{Carbone}(2012)]{carbone12}
{\sc {Carbone}, V.} 2012 {Scalings, Cascade and Intermittency in Solar Wind
  Turbulence}. {\em Space Sci. Rev.\/} {\bf 172}, 343--360.

\bibitem[{Chen} {\em et~al.\/}(2013){Chen}, {Boldyrev}, {Xia} \&
  {Perez}]{chen13}
{\sc {Chen}, C.~H.~K., {Boldyrev}, S., {Xia}, Q. \& {Perez}, J.~C.} 2013
  {Nature of Subproton Scale Turbulence in the Solar Wind}. {\em Phys. Rev.
  Lett.\/} {\bf 110}~(22), 225002.

\bibitem[{Cho}(2011)]{cho11}
{\sc {Cho}, J.} 2011 {Magnetic helicity conservation and inverse energy cascade
  in electron MHD wave packets}. {\em Phys. Rev. Lett.\/} {\bf 106}, 191104.

\bibitem[{Cho} \& {Lazarian}(2004)]{cho04}
{\sc {Cho}, J. \& {Lazarian}, A.} 2004 {The Anisotropy of Electron
  Magnetohydrodynamic Turbulence}. {\em Astrophys. J. Lett.\/} {\bf 615},
  L41--L44.

\bibitem[{Cho} \& {Lazarian}(2009)]{cho09}
{\sc {Cho}, J. \& {Lazarian}, A.} 2009 {Simulations of Electron
  Magnetohydrodynamic Turbulence}. {\em Astrophys. J.\/} {\bf 701}, 236--252.

\bibitem[{Das}(1999)]{das}
{\sc {Das}, A.} 1999 Nonlinear aspects of two-dimensional electron
  magnetohydrodynamics. {\em Plasma Physics and Controlled Fusion\/} {\bf 41},
  A531--A538.

\bibitem[{Das} \& {Diamond}(2000)]{das00}
{\sc {Das}, A. \& {Diamond}, P.~H.} 2000 {Theory of two-dimensional mean field
  electron magnetohydrodynamics}. {\em Phys. Plasmas\/} {\bf 7}, 170--177.

\bibitem[{Diamond} {\em et~al.\/}(2011){Diamond}, {Hasegawa} \&
  {Mima}]{diamond11}
{\sc {Diamond}, P.~H., {Hasegawa}, A. \& {Mima}, K.} 2011 {Vorticity dynamics,
  drift wave turbulence, and zonal flows: a look back and a look ahead}. {\em
  Plasma Physics and Controlled Fusion\/} {\bf 53}~(12), 124001.

\bibitem[{Drake} {\em et~al.\/}(1994){Drake}, {Kleva} \& {Mandt}]{drake}
{\sc {Drake}, J.~F., {Kleva}, R.~G. \& {Mandt}, M.~E.} 1994 {Structure of thin
  current layers: Implications for magnetic reconnection}. {\em Phys. Rev.
  Lett.\/} {\bf 73}, 1251--1254.

\bibitem[{Dreher} {\em et~al.\/}(2005){Dreher}, {Laveder}, {Grauer}, {Passot}
  \& {Sulem}]{dreher}
{\sc {Dreher}, J., {Laveder}, D., {Grauer}, R., {Passot}, T. \& {Sulem}, P.~L.}
  2005 {Formation and disruption of Alfv{\'e}nic filaments in Hall
  magnetohydrodynamics}. {\em Phys. Plasmas\/} {\bf 12}~(5), 052319.

\bibitem[{Frieman} \& {Chen}(1982)]{frieman82}
{\sc {Frieman}, E.~A. \& {Chen}, L.} 1982 {Nonlinear gyrokinetic equations for
  low-frequency electromagnetic waves in general plasma equilibria}. {\em Phys.
  Fluids\/} {\bf 25}, 502--508.

\bibitem[{Galtier}(2003)]{galtier03i}
{\sc {Galtier}, S.} 2003 {Weak inertial-wave turbulence theory}. {\em Phys.
  Rev. E (R)\/} {\bf 68}, 015301.

\bibitem[{Galtier}(2006{\natexlab{{\em a\/}}})]{galtier06a}
{\sc {Galtier}, S.} 2006{\natexlab{{\em a\/}}} {Multi-scale Turbulence in the
  Inner Solar Wind}. {\em J. Low Temp. Physics\/} {\bf 145}, 59--74.

\bibitem[{Galtier}(2006{\natexlab{{\em b\/}}})]{galtier06}
{\sc {Galtier}, S.} 2006{\natexlab{{\em b\/}}} {Wave turbulence in
  incompressible Hall magnetohydrodynamics}. {\em J. Plasma Physics\/} {\bf
  72}, 721--769.

\bibitem[{Galtier}(2008)]{galtier08}
{\sc {Galtier}, S.} 2008 {von K{\'a}rm{\'a}n-Howarth equations for Hall
  magnetohydrodynamic flows}. {\em Phys. Rev. E\/} {\bf 77}~(1), 015302.

\bibitem[{Galtier}(2014)]{galtier14}
{\sc {Galtier}, S.} 2014 {Theory of helical turbulence under fast rotation}.
  {\em Phys. Rev. E (R)\/} {\bf 89}, 041001.

\bibitem[{Galtier} \& {Bhattacharjee}(2003)]{galtier03b}
{\sc {Galtier}, S. \& {Bhattacharjee}, A.} 2003 {Anisotropic weak whistler wave
  turbulence in electron magnetohydrodynamics}. {\em Phys. Plasmas\/} {\bf 10},
  3065--3076.

\bibitem[{Galtier} \& {Bhattacharjee}(2005)]{GB05}
{\sc {Galtier}, S. \& {Bhattacharjee}, A.} 2005 {Anisotropic wave turbulence in
  electron MHD}. {\em Plasma Phys. Controlled Fusion\/} {\bf 47}, B691--B701.

\bibitem[{Galtier} \& {Buchlin}(2007)]{galtier07}
{\sc {Galtier}, S. \& {Buchlin}, E.} 2007 {Multiscale Hall-Magnetohydrodynamic
  Turbulence in the Solar Wind}. {\em Astrophys. J.\/} {\bf 656}, 560--566.

\bibitem[{Galtier} {\em et~al.\/}(2001){Galtier}, {Nazarenko} \&
  {Newell}]{Galtier2001}
{\sc {Galtier}, S., {Nazarenko}, S.~V. \& {Newell}, A.~C.} 2001 {On wave
  turbulence in MHD}. {\em Nonlinear Proc. Geophys.\/} {\bf 8}, 141--150.

\bibitem[{Galtier} {\em et~al.\/}(2005){Galtier}, {Pouquet} \&
  {Mangeney}]{Galtier05}
{\sc {Galtier}, S., {Pouquet}, A. \& {Mangeney}, A.} 2005 {On spectral scaling
  laws for incompressible anisotropic magnetohydrodynamic turbulence}. {\em
  Phys. Plasmas\/} {\bf 12}~(9), 092310.

\bibitem[{Ghosh} \& {Goldstein}(1997)]{Ghosh97}
{\sc {Ghosh}, S. \& {Goldstein}, M.~L.} 1997 {Anisotropy in Hall MHD turbulence
  due to a mean magnetic field}. {\em J. Plasma Phys.\/} {\bf 57}, 129--154.

\bibitem[{Ghosh} {\em et~al.\/}(1996){Ghosh}, {Siregar}, {Roberts} \&
  {Goldstein}]{Ghosh96}
{\sc {Ghosh}, S., {Siregar}, E., {Roberts}, D.~A. \& {Goldstein}, M.~L.} 1996
  {Simulation of high-frequency solar wind power spectra using Hall
  magnetohydrodynamics}. {\em J. Geophys. Res.\/} {\bf 101}, 2493--2504.

\bibitem[{Goldreich} \& {Reisenegger}(1992)]{goldreich92}
{\sc {Goldreich}, P. \& {Reisenegger}, A.} 1992 {Magnetic field decay in
  isolated neutron stars}. {\em Astrophys. J.\/} {\bf 395}, 250--258.

\bibitem[{Goldstein} \& {Roberts}(1999)]{Goldstein99}
{\sc {Goldstein}, M.~L. \& {Roberts}, D.~A.} 1999 {Magnetohydrodynamic
  turbulence in the solar wind}. {\em Phys. Plasmas\/} {\bf 6}, 4154--4160.

\bibitem[{Hasegawa} \& {Chen}(1975)]{hasegawa}
{\sc {Hasegawa}, A. \& {Chen}, L.} 1975 {Kinetic process of plasma heating due
  to Alfv{\'e}n wave excitation}. {\em Phys. Rev. Lett.\/} {\bf 35}, 370--373.

\bibitem[{Hirose} {\em et~al.\/}(2004){Hirose}, {Ito}, {Mahajan} \&
  {Ohsaki}]{hirose}
{\sc {Hirose}, A., {Ito}, A., {Mahajan}, S.~M. \& {Ohsaki}, S.} 2004 {Relation
  between Hall-MHD and the kinetic Alfv{\'e}n wave}. {\em Phys. Lett. A\/} {\bf
  330}, 474--480.

\bibitem[{Hori} \& {Miura}(2008)]{hori}
{\sc {Hori}, D. \& {Miura}, H.} 2008 Spectrum properties of hall mhd
  turbulence. {\em Plasma and Fusion Research\/} {\bf 3}, S1053.

\bibitem[{Howes}(2006)]{howes06}
{\sc {Howes}, G.~G.} 2006 {Limitations of Hall MHD as a model for turbulence in
  weakly collisional plasmas}. {\em Nonlin. Processes Geophys.\/} {\bf 16},
  219--232.

\bibitem[{Howes} {\em et~al.\/}(2008){Howes}, {Cowley}, {Dorland}, {Hammett},
  {Quataert} \& {Schekochihin}]{howes08}
{\sc {Howes}, G.~G., {Cowley}, S.~C., {Dorland}, W., {Hammett}, G.~W.,
  {Quataert}, E. \& {Schekochihin}, A.~A.} 2008 {A model of turbulence in
  magnetized plasmas: Implications for the dissipation range in the solar
  wind}. {\em J. Geophys. Res. (Space Physics)\/} {\bf 113}, 5103.

\bibitem[{Howes} \& {Quataert}(2010)]{howes10}
{\sc {Howes}, G.~G. \& {Quataert}, E.} 2010 {On the Interpretation of Magnetic
  Helicity Signatures in the Dissipation Range Of Solar Wind Turbulence}. {\em
  Astrophys. J. Lett.\/} {\bf 709}, L49--L52.

\bibitem[{Hunana} {\em et~al.\/}(2011){Hunana}, {Laveder}, {Passot}, {Sulem} \&
  {Borgogno}]{hunana11}
{\sc {Hunana}, P., {Laveder}, D., {Passot}, T., {Sulem}, P.~L. \& {Borgogno},
  D.} 2011 {Reduction of Compressibility and Parallel Transfer by Landau
  Damping in Turbulent Magnetized Plasmas}. {\em Astrophys. J.\/} {\bf 743},
  128.

\bibitem[{Karimabadi} {\em et~al.\/}(2013){Karimabadi}, {Roytershteyn}, {Wan},
  {Matthaeus}, {Daughton}, {Wu}, {Shay}, {Loring}, {Borovsky}, {Leonardis},
  {Chapman} \& {Nakamura}]{karimabadi13}
{\sc {Karimabadi}, H., {Roytershteyn}, V., {Wan}, M., {Matthaeus}, W.~H.,
  {Daughton}, W., {Wu}, P., {Shay}, M., {Loring}, B., {Borovsky}, J.,
  {Leonardis}, E., {Chapman}, S.~C. \& {Nakamura}, T.~K.~M.} 2013 {Coherent
  structures, intermittent turbulence, and dissipation in high-temperature
  plasmas}. {\em Phys. Plasmas\/} {\bf 20}~(1), 012303.

\bibitem[{Kingsep} {\em et~al.\/}(1990){Kingsep}, {Chukbar} \&
  {Yankov}]{kingsep}
{\sc {Kingsep}, A.S., {Chukbar}, K.V. \& {Yankov}, V.V.} 1990 {\em Review of
  plasma physics\/}. Consultant bureau, New York, vol. 16.

\bibitem[{Kiyani} {\em et~al.\/}(2009){Kiyani}, {Chapman}, {Khotyaintsev},
  {Dunlop} \& {Sahraoui}]{Kiyani}
{\sc {Kiyani}, K.~H., {Chapman}, S.~C., {Khotyaintsev}, Y.~V., {Dunlop}, M.~W.
  \& {Sahraoui}, F.} 2009 {Global Scale-Invariant Dissipation in Collisionless
  Plasma Turbulence}. {\em Phys. Rev. Lett.\/} {\bf 103}~(7), 075006.

\bibitem[{Lukin}(2009)]{lukin}
{\sc {Lukin}, V.~S.} 2009 {Stationary nontearing inertial scale electron
  magnetohydrodynamic instability}. {\em Phys. Plasmas\/} {\bf 16}~(12),
  122105.

\bibitem[{Lvov} {\em et~al.\/}(2004){Lvov}, {Polzin} \& {Tabak}]{lvov04}
{\sc {Lvov}, Y.V., {Polzin}, K.L. \& {Tabak}, E.G.} 2004 {Energy spectra of the
  ocean's internal wave field: theory and observations}. {\em Phys. Rev.
  Lett.\/} {\bf 92}~(12), 128501.

\bibitem[{Lyutikov}(2013)]{lyutikov}
{\sc {Lyutikov}, M.} 2013 {Electron magnetohydrodynamics: Dynamics and
  turbulence}. {\em Phys. Rev. E\/} {\bf 88}, 053103.

\bibitem[{Mandt} {\em et~al.\/}(1994){Mandt}, {Denton} \& {Drake}]{mandt}
{\sc {Mandt}, M.~E., {Denton}, R.~E. \& {Drake}, J.~F.} 1994 {Transition to
  whistler mediated magnetic reconnection}. {\em Geophys. Res. Lett.\/} {\bf
  21}, 73--76.

\bibitem[{Martin} {\em et~al.\/}(2012){Martin}, {Dmitruk} \& {Gomez}]{martin}
{\sc {Martin}, L.N., {Dmitruk}, P. \& {Gomez}, D.O.} 2012 {Energy spectrum,
  dissipation, and spatial structures in reduced Hall magnetohydrodynamic}.
  {\em Phys. Plasmas\/} {\bf 19}, 052305.

\bibitem[{Matthaeus} \& {Goldstein}(1982)]{bill82}
{\sc {Matthaeus}, W.~H. \& {Goldstein}, M.~L.} 1982 {Measurement of the rugged
  invariants of magnetohydrodynamic turbulence in the solar wind}. {\em J.
  Geophys. Res.\/} {\bf 87}, 6011--6028.

\bibitem[{Matthaeus} {\em et~al.\/}(2014){Matthaeus}, {Oughton}, {Osman},
  {Servidio}, {Wan}, {Gary}, {Shay}, {Valentini}, {Roytershteyn}, {Karimabadi}
  \& {Chapman}]{bill14}
{\sc {Matthaeus}, W.~H., {Oughton}, S., {Osman}, K.~T., {Servidio}, S., {Wan},
  M., {Gary}, S.~P., {Shay}, M.~A., {Valentini}, F., {Roytershteyn}, V.,
  {Karimabadi}, H. \& {Chapman}, S.~C.} 2014 {Nonlinear and Linear Timescales
  near Kinetic Scales in Solar Wind Turbulence}. {\em ArXiv 1404.6569\/} .

\bibitem[{Meyrand} \& {Galtier}(2010)]{meyrand10}
{\sc {Meyrand}, R. \& {Galtier}, S.} 2010 {A Universal Law for Solar-wind
  Turbulence at Electron Scales}. {\em Astrophys. J.\/} {\bf 721}, 1421--1424.

\bibitem[{Meyrand} \& {Galtier}(2012)]{meyrand12}
{\sc {Meyrand}, R. \& {Galtier}, S.} 2012 {Spontaneous Chiral Symmetry Breaking
  of Hall Magnetohydrodynamic Turbulence}. {\em Phys. Rev. Lett.\/} {\bf
  109}~(19), 194501.

\bibitem[{Meyrand} \& {Galtier}(2013)]{meyrand13}
{\sc {Meyrand}, R. \& {Galtier}, S.} 2013 {Anomalous $k_\perp^{-8/3}$ spectrum
  in electron magnetohydrodynamic turbulence}. {\em Phys. Rev. Lett.\/} {\bf
  111}, 264501.

\bibitem[{Mininni} {\em et~al.\/}(2007){Mininni}, {Alexakis} \&
  {Pouquet}]{mininni07}
{\sc {Mininni}, P.~D., {Alexakis}, A. \& {Pouquet}, A.} 2007 {Energy transfer
  in Hall-MHD turbulence: cascades, backscatter, and dynamo action}. {\em J.
  Plasma Physics\/} {\bf 73}, 377--401.

\bibitem[{Nazarenko}(2011)]{nazarenko11}
{\sc {Nazarenko}, S.}, ed. 2011 {\em {Wave Turbulence}\/}, {\em Lecture Notes
  in Physics, Berlin Springer Verlag\/}, vol. 825.

\bibitem[{Ng} {\em et~al.\/}(2003){Ng}, {Bhattacharjee}, {Germaschewski} \&
  {Galtier}]{ng03}
{\sc {Ng}, C.~S., {Bhattacharjee}, A., {Germaschewski}, K. \& {Galtier}, S.}
  2003 {Anisotropic fluid turbulence in the interstellar medium and solar
  wind}. {\em Phys. Plasmas\/} {\bf 10}, 1954--1962.

\bibitem[{Passot} {\em et~al.\/}(2014){Passot}, {Henri}, {Laveder} \&
  {Sulem}]{passot14}
{\sc {Passot}, T., {Henri}, P., {Laveder}, D. \& {Sulem}, P.L.} 2014 {Fluid
  simulations of ion scale plasmas with weakly distorted magnetic field}. {\em
  Submitted\/} {\bf 000}, 000.

\bibitem[{Pouquet} {\em et~al.\/}(1976){Pouquet}, {Frisch} \&
  {Leorat}]{pouquet76}
{\sc {Pouquet}, A., {Frisch}, U. \& {Leorat}, J.} 1976 {Strong MHD helical
  turbulence and the nonlinear dynamo effect}. {\em J. Fluid Mech.\/} {\bf 77},
  321--354.

\bibitem[{Rousculp} \& {Stenzel}(1997)]{rousculp97}
{\sc {Rousculp}, C.~L. \& {Stenzel}, R.~L.} 1997 {Helicity Injection by Knotted
  Antennas into Electron Magnetohydrodynamical Plasmas}. {\em Phys. Rev.
  Lett.\/} {\bf 79}, 837--840.

\bibitem[{Rudakov} {\em et~al.\/}(2011){Rudakov}, {Mithaiwala}, {Ganguli} \&
  {Crabtree}]{rudakov}
{\sc {Rudakov}, L., {Mithaiwala}, M., {Ganguli}, G. \& {Crabtree}, C.} 2011
  {Linear and nonlinear Landau resonance of kinetic Alfv{\'e}n waves:
  Consequences for electron distribution and wave spectrum in the solar wind}.
  {\em Phys. Plasmas\/} {\bf 18}~(1), 012307.

\bibitem[{Sahraoui} {\em et~al.\/}(2012){Sahraoui}, {Belmont} \&
  {Goldstein}]{sahraoui12}
{\sc {Sahraoui}, F., {Belmont}, G. \& {Goldstein}, M.~L.} 2012 {New Insight
  into Short-wavelength Solar Wind Fluctuations from Vlasov Theory}. {\em
  Astrophys. J.\/} {\bf 748}, 100.

\bibitem[{Sahraoui} {\em et~al.\/}(2007){Sahraoui}, {Galtier} \&
  {Belmont}]{Sahraoui07}
{\sc {Sahraoui}, F., {Galtier}, S. \& {Belmont}, G.} 2007 {On waves in
  incompressible Hall magnetohydrodynamics}. {\em J. Plasma Physics\/} {\bf
  73}, 723--730.

\bibitem[{Sahraoui} {\em et~al.\/}(2010){Sahraoui}, {Goldstein}, {Belmont},
  {Canu} \& {Rezeau}]{sahraoui10}
{\sc {Sahraoui}, F., {Goldstein}, M.~L., {Belmont}, G., {Canu}, P. \& {Rezeau},
  L.} 2010 {Three Dimensional Anisotropic k Spectra of Turbulence at Subproton
  Scales in the Solar Wind}. {\em Phys. Rev. Lett.\/} {\bf 105}~(13), 131101.

\bibitem[{Sahraoui} {\em et~al.\/}(2013){Sahraoui}, {Huang}, {Belmont},
  {Goldstein}, {R{\'e}tino}, {Robert} \& {De Patoul}]{fouad2013}
{\sc {Sahraoui}, F., {Huang}, S.~Y., {Belmont}, G., {Goldstein}, M.~L.,
  {R{\'e}tino}, A., {Robert}, P. \& {De Patoul}, J.} 2013 {Scaling of the
  Electron Dissipation Range of Solar Wind Turbulence}. {\em Astrophys. J.\/}
  {\bf 777}, 15.

\bibitem[{Schekochihin} {\em et~al.\/}(2009){Schekochihin}, {Cowley},
  {Dorland}, {Hammett}, {Howes}, {Quataert} \& {Tatsuno}]{sheko09}
{\sc {Schekochihin}, A.~A., {Cowley}, S.~C., {Dorland}, W., {Hammett}, G.~W.,
  {Howes}, G.~G., {Quataert}, E. \& {Tatsuno}, T.} 2009 {Astrophysical
  Gyrokinetics: Kinetic and Fluid Turbulent Cascades in Magnetized Weakly
  Collisional Plasmas}. {\em Astrophys. J. Suppl.\/} {\bf 182}, 310--377.

\bibitem[{Sentoku} {\em et~al.\/}(2003){Sentoku}, {Mima}, {Kaw} \&
  {Nishikawa}]{sentoku}
{\sc {Sentoku}, Y., {Mima}, K., {Kaw}, P. \& {Nishikawa}, K.} 2003 {Anomalous
  Resistivity Resulting from MeV-Electron Transport in Overdense Plasma}. {\em
  Phys. Rev. Lett.\/} {\bf 90}~(15), 155001.

\bibitem[{Shaikh} \& {Zank}(2005)]{shaikh05}
{\sc {Shaikh}, D. \& {Zank}, G.~P.} 2005 {Driven dissipative whistler wave
  turbulence}. {\em Phys. Plasmas\/} {\bf 12}~(12), 122310.

\bibitem[{Shepherd} \& {Cassak}(2010)]{shepherd}
{\sc {Shepherd}, L.~S. \& {Cassak}, P.~A.} 2010 {Comparison of Secondary
  Islands in Collisional Reconnection to Hall Reconnection}. {\em Phys. Rev.
  Lett.\/} {\bf 105}~(1), 015004.

\bibitem[{Smith} {\em et~al.\/}(2006){Smith}, {Hamilton}, {Vasquez} \&
  {Leamon}]{smith06}
{\sc {Smith}, C.~W., {Hamilton}, K., {Vasquez}, B.~J. \& {Leamon}, R.~J.} 2006
  {Dependence of the Dissipation Range Spectrum of Interplanetary Magnetic
  Fluctuations on the Rate of Energy Cascade}. {\em Astrophys. J. Lett.\/} {\bf
  645}, L85--L88.

\bibitem[{Stenzel} \& {Urrutia}(1996)]{stenzel96}
{\sc {Stenzel}, R.~L. \& {Urrutia}, J.~M.} 1996 {Helicity and Transport in
  Electron MHD Heat Pulses}. {\em Phys. Rev. Lett.\/} {\bf 76}, 1469--1472.

\bibitem[{Stenzel} {\em et~al.\/}(1995){Stenzel}, {Urrutia} \&
  {Rousculp}]{stenzel95}
{\sc {Stenzel}, R.~L., {Urrutia}, J.~M. \& {Rousculp}, C.~L.} 1995 {Helicities
  of Electron Magnetohydrodynamic Currents and Fields in Plasmas}. {\em Phys.
  Rev. Lett.\/} {\bf 74}, 702--705.

\bibitem[{Tessein} {\em et~al.\/}(2009){Tessein}, {Smith}, {MacBride},
  {Matthaeus}, {Forman} \& {Borovsky}]{Tessein}
{\sc {Tessein}, J.~A., {Smith}, C.~W., {MacBride}, B.~T., {Matthaeus}, W.~H.,
  {Forman}, M.~A. \& {Borovsky}, J.~E.} 2009 {Spectral Indices for
  Multi-Dimensional Interplanetary Turbulence at 1 AU}. {\em Astrophys. J.\/}
  {\bf 692}, 684--693.

\bibitem[{Voitenko}(1998)]{voitenko98}
{\sc {Voitenko}, Y.~M.} 1998 {Three-wave coupling and weak turbulence of
  kinetic Alfv{\'e}n waves}. {\em J. Plasma Phys.\/} {\bf 60}, 515--527.

\bibitem[{Zakharov} {\em et~al.\/}(1992){Zakharov}, {L'Vov} \&
  {Falkovich}]{ZLF}
{\sc {Zakharov}, V.~E., {L'Vov}, V.~S. \& {Falkovich}, G.} 1992 {\em Kolmogorov
  spectra of turbulence I: Wave turbulence\/}. Springer Series in Nonlinear
  Dynamics, Berlin.

\end{thebibliography}
\end{document}